\documentclass[10pt,graphicx,subfigure,axodraw,cleverref]{article}
\usepackage{amsfonts}
\usepackage{amssymb}
\usepackage[usenames,dvipsnames]{color}
\usepackage{graphicx}
\usepackage{amsmath}
\usepackage{amsfonts}
\usepackage{amssymb}
\usepackage{cleveref}
\usepackage{float}
\usepackage{tabulary}
\usepackage[title]{appendix}
\usepackage{multicol}

\newcommand{\code}[1]{\texttt{#1}}
\newcommand\nocell[1]{\multicolumn{#1}{c|}{}}
\newcommand\nocello[1]{\multicolumn{#1}{|c}{}}

\def\rmuu{\gamma^{\mu}}
\def\rmud{\gamma_{\mu}}
\def\PL{{1-\gamma_5\over 2}}
\def\PR{{1+\gamma_5\over 2}}
\def\sinW2{\sin^2\theta_W}
\def\AEM{\alpha_{EM}}
\def\mul{M_{\tilde{u} L}^2}
\def\mur{M_{\tilde{u} R}^2}
\def\mdl{M_{\tilde{d} L}^2}
\def\mdr{M_{\tilde{d} R}^2}
\def\mz2{M_{z}^2}
\def\c2b{\cos 2\beta}
\def\au{A_u}
\def\ad{A_d}
\def\cob{\cot \beta}
\def\v#1{v_#1}
\def\tb{\tan\beta}
\def\epem{$e^+e^-$}
\def\KK{$K^0$-$\overline{K^0}$}
\def\wi{\omega_i}
\def\xj{\chi_j}
\def\Wmu{W_\mu}
\def\Wnu{W_\nu}
\def\m#1{{\tilde m}_#1}
\def\mH{m_H}
\def\mw#1{{\tilde m}_{\omega #1}}
\def\mx#1{{\tilde m}_{\chi^{0}_#1}}
\def\mc#1{{\tilde m}_{\chi^{+}_#1}}
\def\mwi{{\tilde m}_{\omega i}}
\def\mxi{{\tilde m}_{\chi^{0}_i}}
\def\mci{{\tilde m}_{\chi^{+}_i}}

\def\ch{{\tilde\chi^{+}_1}}
\def\c2{{\tilde\chi^{+}_2}}

\def\tt{{\tilde\theta}}

\def\tp{{\tilde\phi}}

\def\mz{M_z}
\def\sw{\sin\theta_W}
\def\cw{\cos\theta_W}
\def\cb{\cos\beta}
\def\sb{\sin\beta}
\def\rwi{r_{\omega i}}
\def\rxj{r_{\chi j}}
\def\rfp{r_f'}
\def\Kik{K_{ik}}
\def\Fq2{F_{2}(q^2)}
\def\f{\({\cal F}\)}
\def\d1{{\f(\tilde c;\tilde s;\tilde W)+ \f(\tilde c;\tilde \mu;\tilde W)}}
\def\tw{\tan\theta_W}
\def\sec2w{sec^2\theta_W}
\begin{document}
\baselineskip 18pt
\def\today{\ifcase\month\or
 January\or February\or March\or April\or May\or June\or
 July\or August\or September\or October\or November\or December\fi
 \space\number\day, \number\year}
\def\thebibliography#1{\section*{References\markboth
 {References}{References}}\list
 {[\arabic{enumi}]}{\settowidth\labelwidth{[#1]}
 \leftmargin\labelwidth
 \advance\leftmargin\labelsep
 \usecounter{enumi}}
 \def\newblock{\hskip .11em plus .33em minus .07em}
 \sloppy
 \sfcode`\.=1000\relax}
\let\endthebibliography=\endlist
\def\lsim{\ ^<\llap{$_\sim$}\ }
\def\gsim{\ ^>\llap{$_\sim$}\ }
\def\r2{\sqrt 2}
\def\beq{\begin{equation}}
\def\eeq{\end{equation}}
\def\beqn{\begin{eqnarray}}
\def\eeqn{\end{eqnarray}}
\def\rmuu{\gamma^{\mu}}
\def\rmud{\gamma_{\mu}}
\def\PL{{1-\gamma_5\over 2}}
\def\PR{{1+\gamma_5\over 2}}
\def\sinW2{\sin^2\theta_W}
\def\AEM{\alpha_{EM}}
\def\mul{M_{\tilde{u} L}^2}
\def\mur{M_{\tilde{u} R}^2}
\def\mdl{M_{\tilde{d} L}^2}
\def\mdr{M_{\tilde{d} R}^2}
\def\mz2{M_{z}^2}
\def\c2b{\cos 2\beta}
\def\au{A_u}         
\def\ad{A_d}
\def\cob{\cot \beta}
\def\v#1{v_#1}
\def\tb{\tan\beta}
\def\epem{$e^+e^-$}
\def\KK{$K^0$-$\bar{K^0}$}
\def\wi{\omega_i}
\def\xj{\chi_j}
\def\Wmu{W_\mu}
\def\Wnu{W_\nu}
\def\m#1{{\tilde m}_#1}
\def\mH{m_H}
\def\mw#1{{\tilde m}_{\omega #1}}
\def\mx#1{{\tilde m}_{\chi^{0}_#1}}
\def\mc#1{{\tilde m}_{\chi^{+}_#1}}
\def\mwi{{\tilde m}_{\omega i}}
\def\mxi{{\tilde m}_{\chi^{0}_i}}
\def\mci{{\tilde m}_{\chi^{+}_i}}
\def\mz{M_z}
\def\sw{\sin\theta_W}
\def\cw{\cos\theta_W}
\def\cb{\cos\beta}
\def\sb{\sin\beta}
\def\rwi{r_{\omega i}}
\def\rxj{r_{\chi j}}
\def\rfp{r_f'}
\def\Kik{K_{ik}}
\def\Fq2{F_{2}(q^2)}
\def\f{\({\cal F}\)}
\def\d1{{\f(\tilde c;\tilde s;\tilde W)+ \f(\tilde c;\tilde \mu;\tilde W)}}
\def\tw{\tan\theta_W}
\def\sec2w{sec^2\theta_W}
\def\ch{{\tilde\chi^{+}_1}}
\def\c2{{\tilde\chi^{+}_2}}

\def\tt{{\tilde\theta}}

\def\tp{{\tilde\phi}}

\def\mz{M_z}
\def\sw{\sin\theta_W}
\def\cw{\cos\theta_W}
\def\cb{\cos\beta}
\def\sb{\sin\beta}
\def\rwi{r_{\omega i}}
\def\rxj{r_{\chi j}}
\def\rfp{r_f'}
\def\Kik{K_{ik}}
\def\Fq2{F_{2}(q^2)}
\def\f{\({\cal F}\)}
\def\d1{{\f(\tilde c;\tilde s;\tilde W)+ \f(\tilde c;\tilde \mu;\tilde W)}}

\def\b{${\cal{B}}(\mu\to {e} \gamma)$~}
\def\bb{${\cal{B}}(\tau\to {\mu} \gamma)$~}

\def\tw{\tan\theta_W}
\def\sec2w{sec^2\theta_W}
\newcommand{\pn}[1]{{\color{blue}{#1}}}

\newcommand{\aai}[1]{{\color{red}{#1}}}


\begin{titlepage}

\date[\date]
\begin{center}
{\Large {\bf 
Observables of low-lying supersymmetric
 vectorlike leptonic generations via loop corrections
}}\\
\vskip 0.5 true cm
\vspace{1.5cm}
\renewcommand{\thefootnote}
{\fnsymbol{footnote}}
Amin Aboubrahim$^{a}$\footnote{Email: a.abouibrahim@northeastern.edu},
 Tarek Ibrahim$^{b}$\footnote{Email: tibrahim@zewailcity.edu.eg}, Ahmad Itani$^{c}$\footnote{Email: ahmad.itani@liu.edu.lb}
  and Pran Nath$^{a}$\footnote{Email: nath@neu.edu}  
\vskip 0.5 true cm
\end{center}

\begin{center}
{$^a$ Department of Physics, Northeastern University,
Boston, MA 02115-5000, USA} \\
{$^b$ University of Science and Technology, Zewail City of Science and Technology, \\ 6th of October City, Giza 12588, Egypt\footnote{Permanent address:  Department of  Physics, Faculty of Science,
University of Alexandria, Alexandria,  Egypt}\\
}
{$^c$ Department of Physics, Faculty of Science,  Lebanese International University, Beirut, Lebanon} 
\end{center}

\vskip 1.0 true cm

\centerline{\bf Abstract}
A correlated analysis of observables arising from loop induced effects from a vectorlike generation is
given. The observables include flavor changing radiative decays $\mu\to e \gamma, \tau\to \mu \gamma, \tau\to e \gamma$,
electric dipole moments of the charged leptons $e,\mu, \tau$,  and  
corrections to magnetic dipole moments of $g_\mu-2$ and $g_e-2$. 
In this work we give a full analysis of the corrections to these observables 
by taking into account both the supersymmetry loops as well as the exchange of a vectorlike 
leptonic generation. Thus the fermion mass matrix involves a $5\times 5$ mixing matrix while
the scalar sector involves a $10\times 10$ mixing matrix including the CP violating phases from the vectorlike sector.
The analysis is done under the constraint of the Higgs boson mass 
at the experimentally measured value. The loops considered
include the exchange of $W$ and $Z$ bosons and of leptons and a mirror lepton, and the
exchange of charginos and neutralinos, sleptons and mirror sleptons. The correction to the diphoton decay 
of the Higgs $h\to \gamma\gamma$ including the
 exchange of the vectorlike leptonic multiplet is also computed.

\medskip
\end{titlepage}

\section{ Introduction\label{sec:intro}} 

 Precision measurements  can reveal small deviations
  from the standard model (SM) prediction and indicate the existence of new physics beyond the standard model. There are a variety of 
  experiments which are exploring the properties of elementary particles to a high precision to this end. These include 
  flavor changing radiative decays of the charged 
   leptons   $\mu\to e \gamma, \tau\to \mu \gamma$ and $\tau\to e \gamma$, i.e., the MEG experiment~\cite{Adam:2013mnn}, 
 BaBar Collaboration~\cite{Aubert:2009ag} and the Belle Collaboration~\cite{Hayasaka:2007vc},
 the electric dipole moment (EDM) of the electron~\cite{Baron:2016obh}, of the muon  as well as of quarks  ~\cite{Beringer:1900zz}, 
 and the  precision measurement of the anomalous magnetic moment of the muon~\cite{Bennett:2008dy} 
 and of the electron. 
    In this work we explore the implications of a low-lying vectorlike generation on the leptonic processes mentioned above. 
    Vectorlike generations exist in a variety of models including grand unified models, string models
    and D brane models~\cite{vectorlike,Dijkstra:2004cc,Lebedev:2006kn}.   
      Some of these vectorlike generations may be light. Further, vectorlike generations are anomaly free 
    so they preserve good properties of the model as a quantum field theory.    
  The mixings of these light vectorlike generations with the three generations of leptons can lead to 
  contributions to the processes noted above. 
  Several studies of the effects of vectorlike leptons in various processes already exist~\cite{Ibrahim:2010va,Ibrahim:2012ds,Aboubrahim:2013gfa,Ibrahim:2015hva,Martin:2009bg,Graham:2009gy,Martin:2010dc,Martin:2012dg,Moroi:2011aa,Fischler:2013tva,Endo:2011xq}  
   and in non-supersymmetric context in \cite{delAguila:2008pw,ArkaniHamed:2012kq,Kearney:2012zi,Joglekar:2012vc,Ishiwata:2013gma,Ishiwata:2015cga,Ellis:2014dza}.
      In this analysis we perform a correlated study of the contributions of the vectorlike generation to these phenomena. The analysis involves
      an enlarged leptonic mass matrix which is       $5\times 5$
      and a  slepton mass-squared  matrix  which is  $10\times 10$ including the CP violating phases from the vectorlike sector. In the analysis we consider loop exchange of $W$ and $Z$ bosons, leptons and mirror leptons, and exchange of charginos and neutralinos along with the sleptons and mirror sleptons.
      The analysis is done under the constraint of the Higgs boson mass at $\sim 125$ GeV, and an analysis of the contribution to the 
      branching ratio 
       $h\to \gamma\gamma$ from the vectorlike leptonic exchange is also given. \\
The outline of the rest of the paper is as follows: In section~\ref{sec:model} we give a description of the model. 
 In  section~\ref{sec:muegamma} we give an analysis of the
 flavor changing decays of the charged leptons. An analysis of the EDM of the charged leptons is given in  section~\ref{sec:edm}.
  In section~\ref{sec:g-2} we give an analysis of $g-2$ for the charged leptons.  
 An analysis of the contribution of the vectorlike leptonic generation to the diphoton decay of the 
 Higgs boson is given in section ~\ref{sec:h2gg}.
 A numerical analysis is given in section~\ref{sec:numerical} 
  and conclusions are given in section~\ref{sec:conclu}.   
 Further details of the analysis is given in appendices~\ref{sec:appendA} and~\ref{sec:appendB}.

\section{Description of the model \label{sec:model}} 

In this section we give details of the model used in the rest of the paper. As mentioned in  section \ref{sec:intro}
the model consists of three generations of
sequential leptons $(e,\mu,\tau)$ and in addition a single vectorlike generation. Thus one has four sequential families and a mirror
generation. 
The properties of the sequential generation under $SU(3)_C\times SU(2)_L \times U(1)_Y$ are given by
\beqn
\psi_{iL}\equiv \left(
\begin{array}{c}
 \nu_{iL}\\
 \ell_{iL}
\end{array}\right) \sim(1,2,- \frac{1}{2}), \ell^c_{iL}\sim (1,1,1), \nu^c_{iL}\sim (1,1,0),
\eeqn
where the last entry on the right hand side of each $\sim$ is the value of the hypercharge
 $Y$ defined so that $Q=T_3+ Y$ and we have included in our analysis the singlet field
 $\nu^c_i$, where $i$ runs from 1$-$4. The mirrors are given by
\beqn
\chi^c\equiv \left(
\begin{array}{c}
 E_{\mu L}^c\\
 N_L^c
\end{array}\right)
\sim(1,2,\frac{1}{2}), E_{\mu L}\sim (1,1,-1), N_L\sim (1,1,0).
\eeqn
The main difference between the leptons and the mirrors is that while the
 leptons have $V-A$ type interactions with $SU(2)_L\times U(1)_Y$
  gauge bosons the mirrors have  $V+A$ type interactions.

We assume that the mirrors of the vectorlike generation escape acquiring mass at the GUT scale and remain
light down to the electroweak scale
where the superpotential of the model for the lepton part  may be written  in the form
\begin{align}
W&= -\mu \epsilon_{ij} \hat H_1^i \hat H_2^j+\epsilon_{ij}  [f_{1}  \hat H_1^{i} \hat \psi_L ^{j}\hat \tau^c_L
 +f_{1}'  \hat H_2^{j} \hat \psi_L ^{i} \hat \nu^c_{\tau L}
+f_{2}  \hat H_1^{i} \hat \chi^c{^{j}}\hat N_{L}
 +f_{2}'  \hat H_2^{j} \hat \chi^c{^{i}} \hat E_{ L} \nonumber \\
&+ h_{1}   \hat H_1^{i} \hat\psi_{\mu L} ^{j}\hat\mu^c_L
 +h_{1}'  \hat  H_2^{j} \hat\psi_{\mu L} ^{i} \hat\nu^c_{\mu L}
+ h_{2}  \hat  H_1^{i} \hat\psi_{e L} ^{j}\hat e^c_L
 +h_{2}'  \hat  H_2^{j} \hat\psi_{e L} ^{i} \hat\nu^c_{e L}
+y_{5}  \hat  H_1^{i} \hat\psi_{4 L} ^{j} \hat\ell^c_{4 L}
+y_{5} '  \hat  H_2^{j} \hat\psi_{4 L} ^{i} \hat\nu^c_{4 L}
] \nonumber \\
&+ f_{3} \epsilon_{ij}  \hat\chi^c{^{i}}\hat\psi_L^{j}
 + f_{3}' \epsilon_{ij}  \hat\chi^c{^{i}}\hat\psi_{\mu L}^{j}
 + f_{4} \hat\tau^c_L \hat E_{ L}  +  f_{5} \hat\nu^c_{\tau L} \hat N_{L}
 + f_{4}' \hat\mu^c_L \hat E_{ L}  +  f_{5}' \hat\nu^c_{\mu L} \hat N_{L} \nonumber \\
&+ f_{3}'' \epsilon_{ij}  \hat\chi^c{^{i}}\hat\psi_{e L}^{j}
 + f_{4}'' \hat e^c_L \hat E_{ L}  +  f_{5}'' \hat\nu^c_{e L} \hat N_{L}\
+ h_6 \epsilon_{ij}  \hat\chi^c{^{i}}\hat\psi_{4 L}^{j}
+h_7  \hat \ell^c_{4L} \hat E_{ L}
+h_8  \hat \nu^c_{4L} \hat N_{ L}
 ,
 \label{5}
\end{align}
where  $\hat ~$ implies superfields, $\hat\psi_L  \equiv \hat\psi_{\tau L}$ 
stands for $\hat\psi_{3L}$, $\hat\psi_{\mu L}$ stands for $\hat\psi_{2L}$
and  $\hat\psi_{e L}$ stands for $\hat\psi_{1L}$.

The mass terms for the neutrinos, mirror neutrinos,  leptons and  mirror leptons arise from the term
\beq
{\cal{L}}=-\frac{1}{2}\frac{\partial ^2 W}{\partial{A_i}\partial{A_j}}\psi_ i \psi_ j+\text{h.c.},
\label{6}
\eeq
where $\psi$ and $A$ stand for generic two-component fermion and scalar fields.
After spontaneous breaking of the electroweak symmetry, ($\langle H_1^1 \rangle=v_1/\sqrt{2} $ and $\langle H_2^2\rangle=v_2/\sqrt{2}$),
we have the following set of mass terms written in the four-component spinor notation so that
\beq
-{\cal L}_m= \bar\xi_R^T (M_f) \xi_L +\bar\eta_R^T(M_{\ell}) \eta_L +\text{h.c.},
\eeq
where the basis vectors in which the mass matrix is written is given by
\begin{gather}
\bar\xi_R^T= \left(\begin{matrix}\bar \nu_{\tau R} & \bar N_R & \bar \nu_{\mu R}
&\bar \nu_{e R} &\bar \nu_{4 R}\end{matrix}\right),\nonumber\\
\xi_L^T= \left(\begin{matrix} \nu_{\tau L} &  N_L &  \nu_{\mu L}
& \nu_{e L}&\nu_{4 L} \end{matrix}\right),\nonumber\\
\bar\eta_R^T= \left(\begin{matrix}\bar{\tau_ R} & \bar E_R & \bar{\mu_ R}
&\bar{e_ R}
&\bar{\ell}_{4R}
 \end{matrix}\right),\nonumber\\
\eta_L^T= \left(\begin{matrix} {\tau_ L} &  E_L &  {\mu_ L}
& {e_ L}
&\ell_{4L}
\end{matrix}\right),
\end{gather}
and the mass matrix $M_f$ of neutrinos  is given by
\beqn
M_f=
 \left(\begin{matrix} f'_1 v_2/\sqrt{2} & f_5 & 0 & 0 &0\cr
 -f_3 & f_2 v_1/\sqrt{2} & -f_3' & -f_3'' &-h_6\cr
0&f_5'&h_1' v_2/\sqrt{2} & 0 &0\cr
0 & f_5'' & 0 & h_2' v_2/\sqrt{2}&0\cr
0&h_8&0&0&  y_5' v_2/\sqrt{2}
\end{matrix} \right).
\label{7}
\eeqn
We define the matrix elements $(2,2)$ and $(5,5)$ of the mass matrix as $m_N$ and $m^{\nu}_G$, respectively, so that
\beqn
m_N= f_2 v_1/\sqrt 2 ~~~\text{and} ~~~ m^{\nu}_G= y_5' v_2/\sqrt{2}\, .
\label{mn}
\eeqn
The mass matrix is not hermitian and thus one needs biunitary transformations to diagonalize it.
We define the biunitary transformation so that
\beq
D^{\nu \dagger}_R (M_f) D^\nu_L=\text{diag}(m_{\psi_1},m_{\psi_2},m_{\psi_3}, m_{\psi_4}, m_{\psi_5}  ),
\label{Dnu}
\eeq
where
$\psi_1, \psi_2, \psi_3, \psi_4, \psi_5$ are the mass eigenstates for the neutrinos.
In the limit of no mixing
we identify $\psi_1$ as the light tau neutrino, $\psi_2$ as the
heavier mass mirror eigenstate,  $\psi_3$ as the muon neutrino, $\psi_4$ as the electron neutrino and $\psi_5$ as the other heavy four-sequential generation neutrino.
A similar analysis goes to the lepton mass matrix $M_\ell$ where
\beqn
M_\ell=
 \left(\begin{matrix} f_1 v_1/\sqrt{2} & f_4 & 0 & 0 &0\cr
 f_3 & f'_2 v_2/\sqrt{2} & f_3' & f_3'' &h_6\cr
0&f_4'&h_1 v_1/\sqrt{2} & 0 &0\cr
0 & f_4'' & 0 & h_2 v_1/\sqrt{2}&0 \cr
0&h_7&0&0& y_5 v_1/\sqrt{2}
\end{matrix} \right).
\label{8}
\eeqn
We introduce now the mass parameters $m_E$ and $m_G$ for the elements (2,2) and (5,5), respectively, of the mass matrix above so that
\beqn
m_E=  f_2' v_2/\sqrt 2 ~~~ \text{and} ~~~ m_G = y_5 v_1/\sqrt{2}\, .
\label{me}
\eeqn
CP phases that arise from the new sector are defined so that
\beqn
f_i= |f_i|e^{i\chi_i}, ~f'_i= |f'_i|e^{i\chi'_i}, ~f^{''}_i= |f^{''}_i|e^{i\chi^{''}_i}  ~~(i=3,4,5),\nonumber\\
h_k= |h_k|e^{i\chi_k}, ~~k=6,7,8\,.~~~~~~~~~~~~~~~~~~~~~~~~~~~~~~~~~~~~
\label{def-phases}
\eeqn

{
As in the neutrino mass matrix case, 
the charged lepton mass matrix is not hermitian and thus one needs again a biunitary transformation to diagonalize it.
We define the biunitary transformation so that
\beq
D^{\tau \dagger}_R (M_\ell) D^\tau_L=\text{diag}(m_{\tau_1},m_{\tau_2},m_{\tau_3}, m_{\tau_4}, m_{\tau_5}  ),
\label{Dtau}
\eeq
where
$\tau_\alpha$ ($\alpha=$1$-$5)  are the mass eigenstates for the charged lepton matrix. \\
}
The mass-squared matrices of the slepton-mirror slepton and sneutrino-mirror sneutrino  sectors come from three sources: the $F$ term, the $D$ term of the potential and the soft SUSY breaking terms. After spontaneous breaking of the electroweak
symmetry the Lagrangian is given by
\beq
{\cal L}= {\cal L}_F +{\cal L}_D + {\cal L}_{\rm soft},
\eeq
where   $ {\cal L}_F$ is deduced from $-{\cal L}_F=F_i F^*_i$, while the ${\cal L}_D$ is given by
\begin{align}
-{\cal L}_D&=\frac{1}{2} m^2_Z \cos^2\theta_W \cos 2\beta \{\tilde \nu_{\tau L} \tilde \nu^*_{\tau L} -\tilde \tau_L \tilde \tau^*_L
+\tilde \nu_{\mu L} \tilde \nu^*_{\mu L} -\tilde \mu_L \tilde \mu^*_L
+\tilde \nu_{e L} \tilde \nu^*_{e L} -\tilde e_L \tilde e^*_L \nonumber \\
&+\tilde E_R \tilde E^*_R -\tilde N_R \tilde N^*_R
+\tilde \nu_{4 L} \tilde \nu^*_{4 L} -\tilde \ell_{4L} \tilde \ell^*_{4L}
\}
+\frac{1}{2} m^2_Z \sin^2\theta_W \cos 2\beta \{\tilde \nu_{\tau L} \tilde \nu^*_{\tau L}
 +\tilde \tau_L \tilde \tau^*_L
+\tilde \nu_{\mu L} \tilde \nu^*_{\mu L} +\tilde \mu_L \tilde \mu^*_L \nonumber \\
&+\tilde \nu_{e L} \tilde \nu^*_{e L} +\tilde e_L \tilde e^*_L
+\tilde \nu_{4 L} \tilde \nu^*_{4 L}
 +\tilde \ell_{4L} \tilde \ell^*_{4L}\nonumber\\
&-\tilde E_R \tilde E^*_R -\tilde N_R \tilde N^*_R +2 \tilde E_L \tilde E^*_L -2 \tilde \tau_R \tilde \tau^*_R
-2 \tilde \mu_R \tilde \mu^*_R -2 \tilde e_R \tilde e^*_R
-2 \tilde \ell_{4 R} \tilde \ell^*_{4 R}
\},
\label{12}
\end{align}
and ${\cal L}_{\rm soft}$ is given in appendix A.

{

\section{The analysis of $\mathcal{B}(\ell_i \rightarrow \ell_j \gamma)$  with inclusion of 
vectorlike leptons \label{sec:muegamma} }             
Stringent bounds exist on the decay $\mu\to e \gamma$ from the  MEG experiment~\cite{Adam:2013mnn}
\begin{align} 
    {\cal B}(\mu \to e  \gamma) &<5.7 \times 10^{-13} ~~~{\rm at ~} 90\% ~{\rm CL} ~~{\rm (MEG)}.
     \label{1}
\end{align}
 Other flavor changing decays are  $\tau \to \mu\gamma$ and $\tau \to e \gamma$. Here the current experimental limits on the branching ratios of these processes from the BaBar Collaboration~\cite{Aubert:2009ag} and from the Belle Collaboration~\cite{Hayasaka:2007vc} are 
\begin{align} 
{\cal B}(\tau \to \mu  \gamma) &< 4.4 \times 10^{-8} ~~~~{\rm at ~} 90\% ~{\rm CL} ~~{\rm (BaBar)},\nonumber\\
{\cal B}(\tau \to \mu  \gamma) &< 4.5 \times 10^{-8} ~~~~{\rm at ~} 90\% ~{\rm CL} ~~{\rm (Belle)}, \nonumber\\
{\cal B}(\tau \to e  \gamma) &< 3.3 \times 10^{-8} ~~~~{\rm at ~} 90\% ~{\rm CL} ~~{\rm (BaBar)}.   
 \label{2}
 \end{align}
Improvement in the measurements of flavor changing processes is expected  to occur at the SuperB factories~\cite{O'Leary:2010af,Aushev:2010bq,Biagini:2010cc} (for a review see~\cite{Hewett:2012ns}).   
 Thus it is of interest to see if theoretical estimates for these branching ratios can 
  lie close to the current  experimental limits to be detectable in improved experiment. Flavor violating radiative decays have  been analyzed in several previous works (see, e.g., \cite{Hewett:2012ns,Gabbiani:1988rb,Arnowitt:1990ww,Gabbiani:1996hi,Abada:2008ea,Altmannshofer:2009ne,McKeen:2013dma}).
 However, none of these works  explore the class of models discussed here. 

We discuss now the specifics of the model. Thus the decay $\mu \rightarrow e \gamma$ is induced by one-loop electric and magnetic transition dipole moments, which arise from the diagrams of Fig.~\ref{fig3}.
\begin{figure}
\begin{center}
\includegraphics[scale=.35]{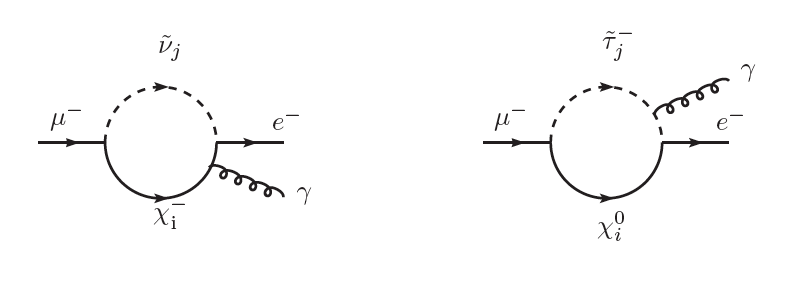}
\includegraphics[scale=.35]{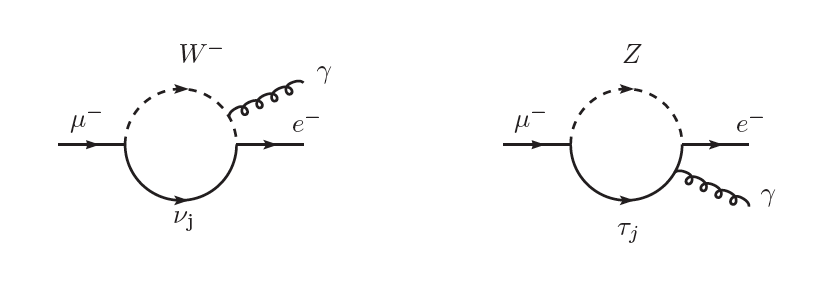}
\caption{The diagrams that allow the decay of $\mu\to e\gamma$ via supersymmetric loops involving the chargino (top left) and the neutralino (top right) and via $W$ loop (bottom left) and $Z$ loop (bottom right) with emission of the photon from the charged particle inside the loop.}
\label{fig3}
\end{center}
\end{figure}
For an incoming muon of momentum $p$ and a resulting electron of momentum $p'$, we define the amplitude
\beq
\langle e (p') | J_{\alpha} | \mu (p)\rangle = \bar{u}_{e} (p') \Gamma_{\alpha} u_{\mu} (p),
\label{32}
\eeq
where
\beq
\Gamma_{\alpha} (q) =\frac{F^{\mu e}_2 (q) i \sigma_{\alpha \beta} q^{\beta}}{m_{\mu} +m_{e}}
+\frac{F^{\mu e}_3 (q)  \sigma_{\alpha \beta} \gamma_5 q^{\beta}}{m_{\mu} +m_{e}}+...,
\label{33}
\eeq
with $q = p' -p$ and where $m_f$ denotes the mass of the fermion $f$.
The branching ratio of $\mu \rightarrow e  \gamma$ is given by
\beqn
   {\cal B} (\mu \rightarrow e  \gamma) =\frac{24 \pi^2}{ G^2_F m^2_{\mu} (m_{\mu}+m_{e})^2} \{|F^{\mu e}_2 (0)|^2
+|F^{\mu e}_3 (0)|^2 \},
\label{34}
\eeqn
where the form factors $F^{\mu e}_2$ and $F^{\mu e}_3$ arise from the chargino, neutralino and vector bosons contributions as follows
\beqn
\label{35}
F^{\mu e}_2 (0) = F^{\mu e}_{2 \chi^+} + F^{\mu e}_{2 \chi^0} + F^{\mu e}_{2 W}+ F^{\mu e}_{2 Z}\,,\\
F^{\mu e}_3 (0) = F^{\mu e}_{3 \chi^+} + F^{\mu e}_{3 \chi^0} + F^{\mu e}_{3 W}+ F^{\mu e}_{3 Z} \,.
\label{36}
\eeqn
It is also useful to define ${\cal B}_m$ and ${\cal B}_e$ as follows
\begin{align}
\label{bm}
{\cal B }_m (\mu \rightarrow e  \gamma) =\frac{24 \pi^2}{ G^2_F m^2_{\mu} (m_{\mu}+m_{e})^2} |F^{\mu e}_2 (0)|^2,\\
{\cal B}_e (\mu \rightarrow e  \gamma) =\frac{24 \pi^2}{ G^2_F m^2_{\mu} (m_{\mu}+m_{e})^2} |F^{\mu e}_3 (0)|^2,
\label{be}
\end{align}
where ${\cal B}_m$ is the branching ratio from the  magnetic dipole operator  and ${\cal B}_e$ is the 
 branching ratio from the electric dipole operator. We discuss now the individual contributions to  $F^{\mu e}_{2}$ and
$F^{\mu e}_{3}$ from supersymmetric and non-supersymmetric loops. \\
The chargino contribution  $F^{\mu e}_{2 \chi^+}$ is given by
\beqn
F^{\mu e}_{2 \chi^+}=\sum_{i=1}^2 \sum_{j=1}^{10} \bigg[\frac{-m_{\mu}(m_{\mu} +m_{e})}{192 \pi^2 m^2_{\tilde{\chi_i}^+}}\{C^L_{4ij} C^{L*}_{3ij} + C^R_{4ij} C^{R*}_{3ij} \} F_4 \left(\frac{M^2_{\tilde{\nu_j}}}{m^2_{\tilde{\chi_i}^+}}\right)\nonumber\\
+ \frac{(m_{\mu} +m_{e})}{64 \pi^2 m_{\tilde{\chi_i}^+}}
\{C^L_{4ij} C^{R*}_{3ij} + C^R_{4ij} C^{L*}_{3ij} \} F_3 \left(\frac{M^2_{\tilde{\nu_j}}}{m^2_{\tilde{\chi_i}^+}}\right)\bigg],
\label{39}
\eeqn
where $F_3(x)$ and $F_4(x)$ are given by
\begin{align}
F_{3}(x)&=\frac{1}{(x-1)^{3}}\left[3x^{2}-4x+1-2x^{2}\ln x \right],
\label{19}
\end{align}
and
\begin{align}
F_{4}(x)&=\frac{1}{(x-1)^{4}}\left[2x^{3}+3x^{2}-6x+1-6x^{2}\ln x \right].
\label{20}
\end{align}

\noindent
The neutralino contribution  $F^{\mu e}_{2 \chi^0}$ is given by
\beqn
F^{\mu e}_{2 \chi^0}= \sum_{i=1}^4 \sum_{j=1}^{10} \bigg[\frac{-m_{\mu}(m_{\mu} +m_{e})}{192 \pi^2 m^2_{\tilde{\chi_i}^0}}
\{C'^L_{4ij} C'^{L*}_{3ij} + C'^R_{4ij} C'^{R*}_{3ij} \} F_2 \left(\frac{M^2_{\tilde{\tau_j}}}{m^2_{\tilde{\chi_i}^0}}\right)\nonumber\\
- \frac{(m_{\mu} +m_{e})}{64 \pi^2 m_{\tilde{\chi_i}^0}}
\{C'^L_{4ij} C'^{R*}_{3ij} + C'^R_{4ij} C'^{L*}_{3ij} \} F_1 \left(\frac{M^2_{\tilde{\tau_j}}}{m^2_{\tilde{\chi_i}^0}}\right)\bigg],
\label{40}
\eeqn
where $F_1(x)$ and $F_2(x)$ are given by 
\begin{align}
F_{1}(x)&=\frac{1}{(x-1)^{3}}\left[1-x^{2}+2x\ln x \right],
\label{22}
\end{align}
and
\begin{align}
F_{2}(x)&=\frac{1}{(x-1)^{4}}\left[-x^{3}+6x^{2}-3x-2-6x\ln x \right].
\label{23}
\end{align}

The contributions  from the $W$ exchange $F^{\mu e}_{2 W}$ is given by  
\beqn
F^{\mu e}_{2 W}= \sum_{i=1}^5 \bigg[\frac{{m_{\mu}(m_{\mu} +m_{e})}}{32 \pi^2 m^2_W}  [C^W_{Li4} C^{W*}_{Li3} 
+C^W_{Ri4} C^{W*}_{Ri3} ] F_W \left(\frac{m^2_{\psi_i}}{m^2_W}\right)\nonumber\\
 + \frac{{m_{\psi_i}(m_{\mu} +m_{e})}}{32 \pi^2 m^2_W}
  { [C^W_{Li4} C^{W*}_{Ri3} 
+C^W_{Ri4} C^{W*}_{Li3} ]} G_W \left(\frac{m^2_{\psi_i}}{m^2_W}\right)\bigg],
\label{41}
\eeqn
where the form factors $F_{W}(x)$ and $G_{W}(x)$ are given by
\begin{align}
F_{W}(x)&=\frac{1}{6(x-1)^{4}}\left[4 x^4- 49x^{3}+18 x^3 \ln x+78x^{2}-43 x +10 \right],
\label{25}
\end{align}
and
\begin{align}
G_{W}(x)&=\frac{1}{(x-1)^{3}}\left[4 -15 x+12 x^2 - x^3-6 x^2 \ln x \right].
\label{26}
\end{align}

The contribution $F^{\mu e}_{2 Z}$ from the $Z$ exchange is given by
\beqn
F^{\mu e}_{2 Z}=  \sum_{\beta=1}^{5} \bigg[\frac{{m_{\mu}(m_{\mu} +m_{e})}}{64 \pi^2 m^2_Z}  [C^Z_{L\beta 4} C^{Z*}_{L\beta 3} 
+C^Z_{R\beta 4} C^{Z*}_{R\beta 3} ]  F_Z \left(\frac{m^2_{\tau_{\beta}}}{m^2_Z}\right)\nonumber\\
 + \frac{{m_{\tau_{\beta}}(m_{\mu} +m_{e})}}{64 \pi^2 m^2_Z} 
  { [C^Z_{L\beta 4} C^{Z*}_{R\beta 3} 
+C^Z_{R\beta 4} C^{Z*}_{L\beta 3} ]} G_Z \left(\frac{m^2_{\tau_{\beta}}}{m^2_Z}\right)\bigg],
\label{42}
\eeqn
where the form factors $F_{Z}(x)$ and $G_{Z}(x)$ are given by
\begin{align}
F_{Z}(x)&=\frac{1}{3(x-1)^{4}}\left[-5 x^4+14x^{3}-39 x^2+18 x^2 \ln x+38 x -8 \right],
\label{28}
\end{align}
and
\begin{align}
G_{Z}(x)&=\frac{2}{(x-1)^{3}}\left[x^3 + 3 x-6 x \ln x-4 \right].
\label{29}
\end{align}

\noindent
The chargino contribution  $F^{\mu e}_{3 \chi^+}$ is given by
\beqn
F^{\mu e}_{3 \chi^+}= \sum_{i=1}^2 \sum_{j=1}^{10} \frac{(m_{\mu} +m_{e})m_{\tilde{\chi_i}^+} }{32 \pi^2 M^2_{\tilde{\nu_j}}}
\left[ C^L_{4ij} C^{R*}_{3ij} - C^R_{4ij} C^{L*}_{3ij} \right] 
F_6 \left(\frac{m^2_{\tilde{\chi_i}^+}}{M^2_{\tilde{\nu_j}}}\right),
\label{43}
\eeqn
where
\beq
F_6(x)= \frac{1}{2(x-1)^2} \left[-x +3 + \frac{2\ln x}{1-x} \right].
\label{44}
\eeq

\noindent
The neutralino contribution  $F^{\mu e}_{3 \chi^0}$ is given by
\beqn
F^{\mu e}_{3 \chi^0}= \sum_{i=1}^4 \sum_{j=1}^{10} \frac{(m_{\mu} +m_{e})m_{\tilde{\chi_i}^0} }{32 \pi^2 M^2_{\tilde{\tau_j}}}
\left[ C'^L_{4ij} C'^{R*}_{3ij} - C'^R_{4ij} C'^{L*}_{3ij} \right] 
F_5 \left(\frac{m^2_{\tilde{\chi_i}^0}}{M^2_{\tilde{\tau_j}}}\right),
\label{38}
\eeqn
where
\beq
F_5(x)=  \frac{1}{2(x-1)^2} \left[x +1 + \frac{2 x \ln x}{1-x} \right].
\label{46}
\eeq

\noindent
The $W$ boson contribution $F^{\mu e}_{3 W}$ is given by
\begin{align}
F^{\mu e}_{3 W}=- \sum_{i=1}^{5} \frac{{m_{\psi_i}(m_{\mu} +m_{e})}}{32 \pi^2 m^2_W}   [C^W_{Li4} C^{W*}_{Ri3} 
-C^W_{Ri4} C^{W*}_{Li3} ] 
I_1\left(\frac{m^{2}_{{\psi}_{i}}}{m^{2}_{W}}\right),
\label{47}
\end{align}
where the form factor $I_1$  is given by
\begin{align}
I_1(x)&=\frac{2}{(1-x)^{2}}\left[1-\frac{11}{4}x +\frac{1}{4}x^2-\frac{3 x^2\ln x}{2(1-x)} \right].
\label{48}
\end{align}
And finally,  the $Z$ exchange diagram contribution $F^{\mu e}_{3 Z}$  is given by
\begin{align}
F^{\mu e}_{3 Z}=   \sum_{\beta=1}^{5}\frac{{(m_{\mu} +m_{e})}}{32 \pi^2}  \frac{m_{\tau_\beta}}{m^2_Z}
[C^Z_{L4 \beta } C^{Z*}_{R3 \beta } 
-C^Z_{R4 \beta } C^{Z*}_{L3 \beta } ] 
I_2\left(\frac{m^{2}_{\tau_{\beta}}}{m^{2}_{Z}}\right),
\label{49}
\end{align}
where the form factor $I_2$  is given by
\begin{align}
I_2(x)&=\frac{2}{(1-x)^{2}}\left[1+\frac{1}{4}x +\frac{1}{4}x^2+\frac{3 x\ln x}{2(1-x)} \right]\,.
\label{50}
\end{align}
}
All couplings $C^L, C^R$, $C^{'L}, C^{'R}$, $C^W_L, C^W_R$, $C^Z_L$ and $C^Z_R$ in Eqs.~(\ref{39})$-$(\ref{43}), (\ref{38}), (\ref{47}) and (\ref{49}), are given in appendix~\ref{sec:appendB}.\\   
An analysis for ${\cal B} (\tau \rightarrow e  \gamma)$ can be done similarly so that
\beqn
   {\cal B} (\tau \rightarrow e  \gamma) =\frac{24 \pi^2}{ G^2_F m^2_{\tau} (m_{\tau}+m_{e})^2} \{|F^{\tau e}_2 (0)|^2
+|F^{\tau e}_3 (0)|^2 \},
\label{34a}
\eeqn
where
 the expressions for  the form factors, $F^{\tau e}_2$ and $F^{\tau e}_3$, can be obtained from Eqs.~(\ref{35}) and (\ref{36}) by 
  the replacements: $m_{\mu}\rightarrow m_{\tau}$ and $C_{3ij}, C'_{3ij},C^W_{i3},C^Z_{\beta 3}\longrightarrow C_{1ij}, C'_{1ij},C^W_{i1},C^Z_{\beta 1}$. \\
Also for ${\cal B} (\tau \rightarrow \mu \gamma)$ we have
\beqn
   {\cal B} (\tau \rightarrow \mu  \gamma) =\frac{24 \pi^2}{ G^2_F m^2_{\tau} (m_{\tau}+m_{\mu})^2} \{|F^{\tau\mu}_2 (0)|^2
+|F^{\tau\mu}_3 (0)|^2 \},
\label{34b}
\eeqn 
where the expressions for  the form factors $F^{\tau\mu}_2$ and $F^{\tau\mu}_3$ can be deduced from Eqs.~(\ref{35}) and (\ref{36}) by the replacements: $m_{\mu}\rightarrow m_{\tau}$, $m_e\rightarrow m_{\mu}$, $C_{3ij}, C'_{3ij},C^W_{i3},C^Z_{\beta 3}\longrightarrow C_{1ij}, C'_{1ij},C^W_{i1},C^Z_{\beta 1}$ and $C_{4ij}, C'_{4ij},C^W_{i4},C^Z_{\beta 4}\longrightarrow C_{3ij}, C'_{3ij},C^W_{i3},C^Z_{\beta 3}$.

{
\section{EDM analysis by inclusion of vectorlike leptons\label{sec:edm}}

The electric dipole moment (EDM) of elementary particles arises only at the multi-loop level in the standard model 
and is beyond the scope of observation in the current or in the near future experiment. However, beyond the standard
model physics can generate EDMs which are within the range of observability.  The current experimental 
limits on $d_e, d_\mu, d_\tau$ are as follows. For the electron we have~\cite{Baron:2016obh}
\begin{align}
d_e <9.3 \times 10^{-29}  e{\rm cm} ~~~(90\% ~\rm CL).
\label{de}
\end{align}
For the muon the current limit on the EDM is~\cite{Bennett:2008dy}
\begin{align}
d_\mu <1.9 \times 10^{-19}  e{\rm cm} ~~~(95\% ~\rm CL).
\label{dm}
\end{align}
The current experimental limit on the EDM of the tau lepton is~\cite{Escribano:1996wp}
\begin{align}
d_\tau <  1.1 \times 10^{-17} e{\rm cm}.
\label{dt}
\end{align}
Next we discuss the case when we include a vectorlike leptonic multiplet which mixes with the three
generations of leptons. In this case the mass eigenstates will be linear combinations of the three
generations plus the vectorlike generation which includes mirror particles. Here we discuss the contribution of the 
model to the lepton EDM. These contributions arise from four sources:  the chargino exchange, the neutralino exchange,
the  $W$ boson exchange and the $Z$ boson exchange. 

\begin{figure}[t]
\begin{center}
{\rotatebox{0}{\resizebox*{12cm}{!}{\includegraphics{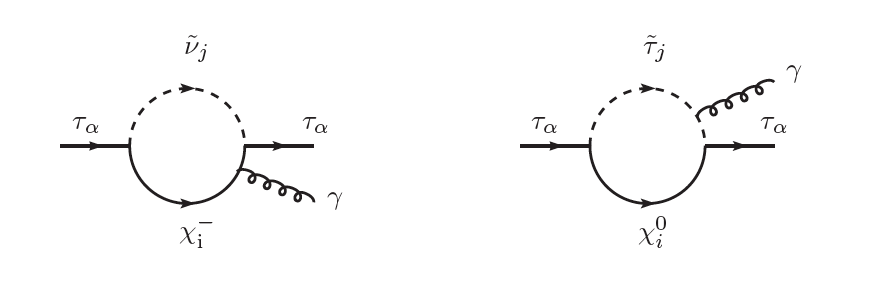}}\hglue5mm}}\\
{\rotatebox{0}{\resizebox*{12cm}{!}{\includegraphics{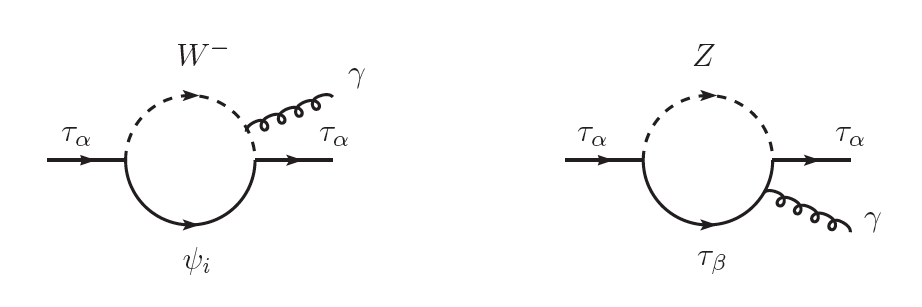}}\hglue5mm}}
\caption{
Upper diagrams: Supersymmetric contributions to the leptonic EDMs
arising from the exchange of the charginos, sneutrinos
and mirror sneutrinos (upper left) and the exchange of neutralinos, sleptons, and mirror sleptons (upper right) 
inside the loop. Lower diagrams: 
Non-supersymmetric  diagrams that contribute to the leptonic EDMs
via the exchange of  the $W$, the sequential and vector like neutrinos (lower left) and the exchange of the 
$Z$, the sequential and vector like charged leptons (lower right).}
 \label{fig4}
\end{center}
\end{figure}
Using the interactions given in appendix~\ref{sec:appendB}, the chargino contribution  is given by
\begin{align}
d_{\alpha}^{\chi^{+}}&=-\frac{1}{16\pi^2}\sum_{i=1}^{2}\sum_{j=1}^{10}\frac{m_{\chi^{+}_i}}{m^2_{\tilde\nu_{j}}}\text{Im}(C^{L}_{\alpha ij}C^{R*}_{\alpha ij})
F_6\left(\frac{m^{2}_{{\chi^{+}_i}}}{m^{2}_{\tilde\nu_{j}}}\right), 
\label{3.1}
\end{align}
where the form factor $F_6(x)$  is given by Eq.~({\ref{44}}).\\ 
Using the interactions given in  appendix B, the neutralino contribution  is given by
\begin{align}
d_{\alpha}^{\chi^{0}}&=-\frac{1}{16\pi^2}\sum_{i=1}^{4}\sum_{j=1}^{10}\frac{m_{\chi^{0}_i}}{m^2_{\tilde\tau_{j}}}\text{Im}(C'^{L}_{\alpha ij}C'^{R*}_{\alpha ij})
F_5\left(\frac{m^{2}_{{\chi^{0}_i}}}{m^{2}_{\tilde\tau_{j}}}\right), 
\end{align}
where the form factor $F_5(x)$  is given by Eq.~({\ref{46}}). \\
 The contributions to the lepton electric dipole moment from the $W$ and $Z$ exchange arise from similar loops.  Using the interactions given in appendix~\ref{sec:appendB} the contribution arising from the $W$ exchange diagram is given by
\begin{align}
d_{\alpha}^{W}&=\frac{1}{16\pi^2}\sum_{i=1}^{5}\frac{m_{\psi^{+}_i}}{m^2_W}\text{Im}(C^{W}_{Li\alpha }C^{W*}_{R i\alpha })
I_1\left(\frac{m^{2}_{{\psi}_{i}}}{m^{2}_{W}}\right),
\end{align}
where the form factor $I_1$  is given by  Eq.~({\ref{48}}). \\

The $Z$ boson exchange diagram contribution is given by
\begin{align}
d_{\alpha}^{Z}&=-\frac{1}{16\pi^2}\sum_{\beta=1}^{5}\frac{m_{\tau_\beta}}{m^2_Z}\text{Im}(C^{Z}_{L\alpha\beta }C^{Z*}_{R \alpha\beta })
I_2\left(\frac{m^{2}_{\tau_{\beta}}}{m^{2}_{Z}}\right), 
\end{align}
where the form factor $I_2$  is given by Eq.~({\ref{50}}). 
Again, all couplings $C^L, C^R$, $C^{'L}, C^{'R}$, $C^W_L, C^W_R$, $C^Z_L$ and $C^Z_R$ used here are given in appendix~\ref{sec:appendB}.


\section{Analysis of $g-2$   with exchange of vectorlike leptons\label{sec:g-2}}
The current experimental result  for the muon $g-2$~\cite{Beringer:1900zz} is 
\begin{align}
\Delta a_{\mu} =a_{\mu}^{\rm exp}- a_{\mu}^{\rm SM}=(28.8 \pm 7.9)\times 10^{-10},
\label{eq1}
\end{align}
which is about  a three sigma deviation from the standard model prediction.  
For  the electron  $g_e-2$ experiment gives~\cite{Patrignani:2016xqp}
\begin{align}
 \Delta a_e= a_e^{\rm exp}- a_e^{\rm SM}= 8.70 (8.07) \times  10^{-13}.
 \label{eq2}
\end{align}
This result relies on a QED calculation up to four loops.
Thus along with Eq.~(\ref{eq1}), Eq.~(\ref{eq2})  also acts as a constraint on the standard model
extensions. We compute beyond the standard model contributions to these within the model of section \ref{sec:model}.
Below we discuss details of the various contributions. The contribution
arising from the exchange of the charginos,  sneutrinos and mirror sneutrinos  as shown in the left diagram in Fig.~\ref{fig1} is
given by
\begin{align}
a_{\alpha}^{\chi^{+}}&=-\sum_{i=1}^{2}\sum_{j=1}^{10}\frac{m_{\tau_{\alpha}}}{16\pi^{2}m_{\chi_{i}^{-}}}\text{Re}(C^{L}_{\alpha ij}C^{R*}_{\alpha ij})
F_{3}\left(\frac{m^{2}_{\tilde{\nu}_{j}}}{m^{2}_{\chi^{-}_{i}}}\right) \nonumber \\
&+\sum_{i=1}^{2}\sum_{j=1}^{10}\frac{m^{2}_{\tau_{\alpha}}}{96\pi^{2}m^{2}_{\chi_{i}^{-}}}\left[|C^{L}_{\alpha ij}|^{2}+|C^{R}_{\alpha ij}|^{2}\right]
F_{4}\left(\frac{m^{2}_{\tilde{\nu}_{j}}}{m^{2}_{\chi^{-}_{i}}}\right),
\label{3}
\end{align}
{where $m_{\chi_{i}^{-}}$ is the mass of chargino $\chi_{i}^{-}$ and $m_{\tilde{\nu}_{j}}$  is the mass of 
sneutrino ${\tilde{\nu}_{j}}$} and where the form factors $F_3$ and $F_4$ are given by Eqs.~(\ref{19}) and (\ref{20}).

\begin{figure}[t]
\begin{center}
{\rotatebox{0}{\resizebox*{10cm}{!}{\includegraphics{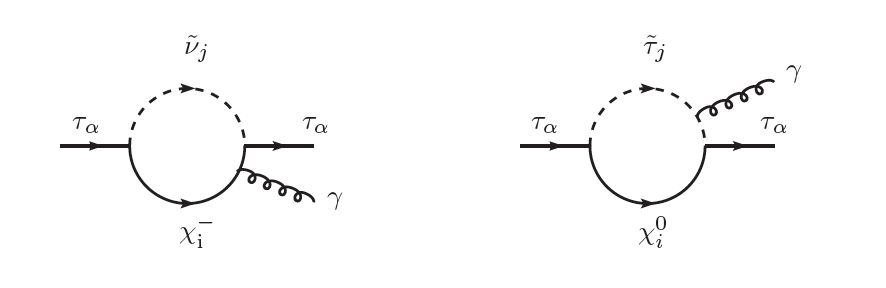}}\hglue5mm}}
\caption{The diagrams that contribute to the leptonic ($\tau_{\alpha}$)
magnetic dipole moment via
  exchange of charginos ($\chi_i^{-}$), sneutrinos and mirror sneutrinos  ($\tilde \nu_j$)   (left diagram) inside the loop and from the exchange
  of neutralinos ($\chi_i^0$),   sleptons   and mirror sleptons ($\tilde \tau_j$) (right diagram) inside the loop.}
\label{fig1}
\end{center}
\end{figure}

\begin{figure}[t]
\begin{center}
{\rotatebox{0}{\resizebox*{10cm}{!}{\includegraphics{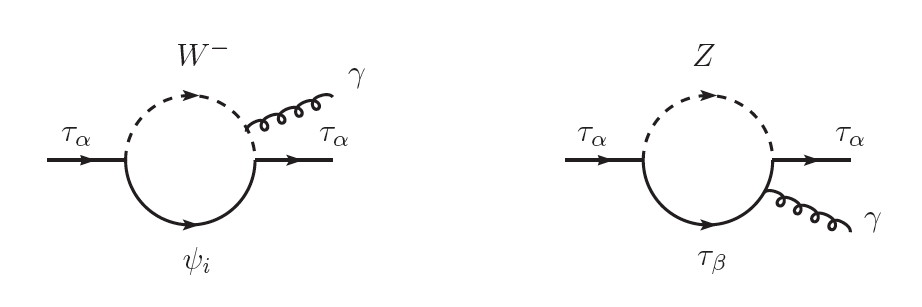}}\hglue5mm}}
\caption{ The W loop  (the left diagram) involving the exchange of sequential and vectorlike neutrinos $\psi_i$
and the Z loop (the right diagram) involving the exchange of sequential and vectorlike charged leptons $\tau_{\beta}$
that contribute to the magnetic dipole moment of the charged lepton $\tau_{\alpha}$.}
\label{fig2}
\end{center}
\end{figure}

The contribution  arising from the exchange of neutralinos, charged sleptons and charged mirror sleptons as shown in the right diagram of Fig.~\ref{fig1} is given by
\begin{align}
a_{\alpha}^{\chi^{0}}&=\sum_{i=1}^{4}\sum_{j=1}^{10}\frac{m_{\tau_{\alpha}}}{16\pi^{2}m_{\chi_{i}^{0}}}\text{Re}(C^{'L}_{\alpha ij}C^{'R*}_{\alpha ij})
F_{1}\left(\frac{m^{2}_{\tilde{\tau}_{j}}}{m^{2}_{\chi^{0}_{i}}}\right) \nonumber \\
&+\sum_{i=1}^{4}\sum_{j=1}^{10}\frac{m^{2}_{\tau_{\alpha}}}{96\pi^{2}m^{2}_{\chi_{i}^{0}}}\left[|C^{'L}_{\alpha ij}|^{2}+|C^{'R}_{\alpha ij}|^{2}\right]
F_{2}\left(\frac{m^{2}_{\tilde{\tau}_{j}}}{m^{2}_{\chi^{0}_{i}}}\right),
\label{10}
\end{align}
where the form factors $F_1$ and $F_2$ are given by Eqs.~(\ref{22}) and (\ref{23}). \\
Next we compute the contribution from the exchange of the $W$ and $Z$ bosons.
Thus the exchange of the $W$ and the exchange of neutrinos and mirror neutrinos as shown
in the left diagram of Fig.~\ref{fig2} gives
\beqn
a^W_{\tau_{\alpha}} = \frac{m^2_{\tau_{\alpha}}}{16 \pi^2 m^2_W} \sum_{i=1}^{5} \left[ [|C^W_{Li\alpha}|^2
+|C^W_{Ri\alpha}|^2] F_W \left(\frac{m^2_{\psi_i}}{m^2_W}\right) + \frac{m_{\psi_i}}{m_{\tau_{\alpha}}} \text{Re}(C^W_{L i \alpha}C^{W*}_{R i \alpha}) G_W \left(\frac{m^2_{\psi_i}}{m^2_W}\right)\right],
\label{24}
\eeqn
where the form factors $F_W$ and $G_W$ are given by Eqs.~(\ref{25}) and (\ref{26}). \\
Finally the exchange of the $Z$ and the exchange of leptons  and mirror leptons  as shown in the right diagram of Fig.~\ref{fig2} gives
\beqn
a^Z_{\tau_{\alpha}} = \frac{m^2_{\tau_{\alpha}}}{32 \pi^2 m^2_Z} \sum_{\beta=1}^{5} \left[ [|C^Z_{L \beta \alpha}|^2
+|C^Z_{R\beta \alpha}|^2] F_Z \left(\frac{m^2_{\tau_{\beta}}}{m^2_Z}\right) + \frac{m_{\tau_{\beta}}}{m_{\tau_{\alpha}}} \text{Re}(C^Z_{L \beta \alpha}C^{Z*}_{R \beta \alpha}) G_Z \left(\frac{m^2_{\tau_{\beta}}}{m^2_Z}\right)\right],
\label{29z}
\eeqn
where the form factors $F_Z$ and $G_Z$ are given by Eqs.~(\ref{28}) and (\ref{29}) and $m_Z$  is the $Z$ boson mass. 
The couplings that enter in Eqs.~(\ref{3}),~(\ref{10}),~(\ref{24}) and~(\ref{29z}) are given in appendix~\ref{sec:appendB}.
For other works relating the muon anomalous magnetic moment to new physics see~\cite{Lindner:2016bgg,Crivellin:2018qmi}.

\section{Leptonic vectorlike contribution to $h\rightarrow\gamma\gamma$\label{sec:h2gg}}

The observed diphoton decay of the Higgs boson shows an agreement with the standard model prediction within the limits of uncertainty which is still significant. As more data is collected and uncertainties better modeled, the signal strength, $R_{\gamma\gamma}$, will be measured with a larger accuracy and any new physics manifest as particles in the loop will be better probed. Thus the ATLAS and CMS collaborations~\cite{Aaboud:2018xdt,Sirunyan:2018ouh} report a signal strength of
\begin{equation}
R_{\gamma\gamma}\equiv\frac{\sigma(pp\rightarrow h)_{\rm obs}}{\sigma(pp\rightarrow h)_{\rm SM}}\cdot\frac{\Gamma(h\rightarrow\gamma\gamma)_{\rm obs}}{\Gamma(h\rightarrow\gamma\gamma)_{\rm SM}}=0.99^{+0.15}_{-0.14} ~~\rm (ATLAS), ~1.18^{+0.17}_{-0.14} ~~\rm (CMS).
\label{Rgglimits}
\end{equation}
In the SM, the largest contribution to $h\rightarrow \gamma\gamma$ comes from the exchange of $W$ bosons and top quarks in the loop.
Thus the SM decay width of a Higgs boson of mass $m_h$ may be approximated by the expression~\cite{Feng:2013mea}
\begin{equation}
\Gamma_{\rm SM}(h\rightarrow \gamma\gamma)\approx \frac{\alpha^2_{\rm em}m^3_h}{256v^2\pi^3}|A_1(\tau_W)+N_c Q_t^2A_{\frac{1}{2}}(\tau_t)|^2\rightarrow \frac{\alpha^2_{\rm em}m^3_h}{256v^2\pi^3}|\mathcal{A}_{\rm SM}|^2,
\end{equation}
where $\mathcal{A}_{\rm SM}\approx -6.49$, $A_1$ and $A_{\frac{1}{2}}$ are loop functions (see Appendix of~\cite{Feng:2013mea}), $\tau_i=4m_i^2/m^2_h$, $N_c$ is the color number and $Q_t$ the top quark charge. The inclusion of SUSY allows for the exchange of heavier particles in the loop. 
In general the decay width  of $h\to \gamma\gamma$ in supersymmetry takes the form
\begin{align}
\Gamma_{\rm SUSY}(h\rightarrow \gamma\gamma)&\approx \frac{\alpha^2_{\rm em}m^3_h}{256v^2\pi^3}\Bigg|\sin(\beta-\alpha)Q^2_W A_1(\tau_W)+\frac{\cos\alpha}{\sin\beta}N_t Q^2_t A_{\frac{1}{2}}(\tau_t) \nonumber \\
&+\frac{b_{\frac{1}{2}}v}{2}N_f Q_f^2\left(\cos\alpha\frac{\partial}{\partial v_2}\log m^2_f-\sin\alpha\frac{\partial}{\partial v_1}\log m^2_f\right) \nonumber \\
&+\frac{b_{0}v}{2}N_{c,S} Q_S^2\left(\cos\alpha\frac{\partial}{\partial v_2}\log m^2_S-\sin\alpha\frac{\partial}{\partial v_1}\log m^2_S\right)\Bigg|^2,
\end{align}
where $\alpha$ is the CP-even Higgs mixing angle, $Q_W$ is the $W$-boson charge, $b_{\frac{1}{2}}=\frac{4}{3}$ (for Dirac fermions of mass $m_f$, number $N_f$ and charge $Q_f$) and $b_0=\frac{1}{3}$ (for charged scalars of mass $m_S$, number $N_{c,S}$ and charge $Q_S$). \\
The inclusion of the vectorlike leptonic generation contributes to the fermionic and scalar parts where the latter is due to the supersymmetric partners of the vectorlike leptons. \\
In this analysis the couplings of the Higgs boson to the first three generations are assumed negligible in comparison with the vectorlike counterparts. Hence the mixings between the vectorlike generation and the first three generations in Eq.~(\ref{7}) can be assumed negligible and so the lepton mass matrix from the vectorlike generation may be written as
\beqn
M^v_f=
 \left(\begin{matrix} f'_2 v_2/\sqrt{2} & h_6\cr
 h_7 & y_5 v_1/\sqrt{2}
\end{matrix} \right).
\label{mvf}
\eeqn
The two mass-squared eigenvalues resulting from diagonalizing the matrix of Eq.~(\ref{mvf}) are
\begin{align}
m^2_{1,2}&=\frac{1}{4}\Big[2 |h_6|^2+2 |h_7|^2+y_5^2 v_1^2+f^{'2}_2 v_2^2 \nonumber \\
&\pm \sqrt{(2 |h_6|^2+2 |h_7|^2+y_5^2 v_1^2+f^{'2}_2 v_2^2)^2-4|2h_6h_7-f'_2 y_5 v_1 v_2|^2}\Big].
\end{align}
Calculating the vectorlike fermionic contribution, one finds that
\begin{equation}
\sum_i\left[\cos\alpha\frac{\partial}{\partial v_2}\log m^2_i-\sin\alpha\frac{\partial}{\partial v_1}\log m^2_i\right]=-\frac{f'_2 y_5 v}{m_1m_2}\cos(\alpha+\beta).
\end{equation}
Considering only this fermionic contribution, we find that the Higgs diphoton rate is enhanced by a factor of
\begin{align}
\frac{\Gamma(h\rightarrow\gamma\gamma)}{\Gamma(h\rightarrow\gamma\gamma)_{\rm SM}}&\approx \Big|1+\frac{1}{\mathcal{A}_{\rm SM}}b_{\frac{1}{2}}N_f Q^2_f\frac{-v^2 f'_2 y_5}{2m_1m_2}\cos(\alpha+\beta)\Big|^2 \nonumber \\
&\approx\Big|1+0.1N_f \frac{v^2 f'_2 y_5}{m_1m_2}\cos(\alpha+\beta)\Big|^2\equiv |1+r_f|^2.
\label{rf}
\end{align}
Now turning to the bosonic contribution which is due to the four scalar superpartners of the vectorlike leptons. The mass eigenvalues are obtained from a $4\times 4$ mass-squared mixing matrix and in the basis $(\tilde E_L, \tilde E_R, \tilde\ell_{4L}, \tilde\ell_{4R})$ is given by
\beqn
 \frac{1}{\sqrt{2}}
\left(\begin{array}{@{}c|c@{}}
\sqrt{2}(M^2_{\tilde E})_{2\times 2} &
  \begin{matrix}
  f'_2 v_2 h_6+y_5 v_1 h^*_7 & 0 \\
  0 & f'_2 v_2 h^*_7+y_5 v_1 h_6
  \end{matrix} \\
  \hline
  \begin{matrix}
  f'_2 v_2 h^*_6+y_5 v_1 h_7 & 0 \\
  0 & f'_2 v_2 h_7+y_5 v_1 h^*_6
  \end{matrix}
  & \sqrt{2}(M^2_{\tilde\ell_{4}})_{2\times 2}  \\
\end{array}\right)_{4\times 4},
\label{mvs}
\eeqn
where $(M^2_{\tilde\ell_{4}})_{2\times 2}$ is given by
\beqn
(M^2_{\tilde\ell_{4}})_{2\times 2}=
 \left(\begin{matrix}  \tilde M^2_{4 L} +\frac{v^2_1|y_5|^2}{2} +|h_6|^2 -m^2_Z \cos 2 \beta \left(\frac{1}{2}-\sin^2\theta_W\right) & \frac{1}{\sqrt{2}}y_5(A^*_{4\ell}v_1-\mu v_2)  \cr
\frac{1}{\sqrt{2}}y_5(A_{4\ell}v_1-\mu^* v_2)   & \tilde M^2_{4} +\frac{v^2_1|y_5|^2}{2} +|h_7|^2 -m^2_Z \cos 2 \beta \sin^2\theta_W
\end{matrix} \right),
\eeqn
and $(M^2_{\tilde E})_{2\times 2}$ is given by
\beqn
(M^2_{\tilde E})_{2\times 2}=
 \left(\begin{matrix}  \tilde M^2_{\chi} +\frac{v^2_2|f'_2|^2}{2}+|h_6|^2+m^2_Z \cos 2 \beta \left(\frac{1}{2}-\sin^2\theta_W\right) & \frac{1}{\sqrt{2}}f'_2(A_E^* v_2-\mu v_1) \cr
\frac{1}{\sqrt{2}}f'_2(A_E v_2-\mu^* v_1)  & \tilde M^2_E +\frac{v^2_2|f'_2|^2}{2}+|h_7|^2+m^2_Z \cos 2 \beta\sin^2\theta_W
\end{matrix} \right).
\eeqn
In this analysis, the scalar masses-squared, $\tilde M^2_{4 L}, \tilde M^2_{4},\tilde M^2_{\chi},\tilde M^2_E$ are much larger than the vectorlike masses, $|h_6|, |h_7|$ and so the $4\times 4$ mass-squared matrix becomes block diagonal. Thus the two mass-squared matrices are now decoupled with superpartner $\tilde\ell_{4_{1,2}}$ for the first and $\tilde E_{1,2}$ for the second. The total bosonic contribution is the sum of the contributions coming from the two decoupled mass-squared matrices and can be written as
\begin{equation}
r_b=r_1+r_2\equiv\frac{1}{\mathcal{A}_{\rm SM}}\frac{b_0 v}{2}Q^2_S   (\Sigma_1+\Sigma_2).
\label{rb}
\end{equation}
Here
\begin{equation}
\Sigma_1=\frac{1}{m^2_{\tilde\ell_{4_1}}m^2_{\tilde\ell_{4_2}}}\left[\left(2\sin^2\theta_W M^2_{11}+\cos2\theta_W M^2_{22}\right)\frac{m^2_Z}{v}\sin(\alpha+\beta)+\sqrt{2}M^2_{12}y_5(A_{4L}\sin\alpha+\mu\cos\alpha)\right],
\end{equation}
and
\begin{equation}
\Sigma_2=\frac{1}{m^2_{\tilde E_1}m^2_{\tilde E_2}}\left[\left(-2\sin^2\theta_W M'^2_{11}-\cos2\theta_W M'^2_{22}\right)\frac{m^2_Z}{v}\sin(\alpha+\beta)-\sqrt{2}M'^2_{12}f'_2(A_E \cos\alpha+\mu\sin\alpha)\right],
\end{equation}
where, for convenience, we renamed the matrices as $M^2\equiv M^2_{\tilde\ell_4}$ and $M'^2\equiv M^2_{\tilde E}$.
Assuming 
${\sigma(pp\rightarrow h)_{\rm obs}}={\sigma(pp\rightarrow h)_{\rm SM}}$ the  enhancement factor $R_{\gamma\gamma}$ is given by
\begin{equation}
R_{\gamma\gamma}=|1+r_f+r_b|^2.
\end{equation}


\section{Numerical Analysis\label{sec:numerical}}

Here we present a correlated analysis of the observables discussed in the previous sections including the effect of vectorlike leptons
(for other works  related to vectorlike  leptons see \cite{other-vectorlike,Aboubrahim:2016xuz}).
In the analysis we will include
the CP violating phases from the vectorlike generation. SUSY CP phases are known to affect electroweak phenomena and these effects can be very significant~\cite{Nath:1991dn,Kizukuri:1992nj,Ibrahim:1998je,Ibrahim:1997gj,Falk:1998pu,Brhlik:1998zn,Ibrahim:1999af,Ibrahim:2011im,Gomez:2006uv,Ibrahim:2007fb}.
In the analysis we use SUGRA model~\cite{Chamseddine:1982jx} with
 non-universal soft parameters given by
$m_0, ~A_0, ~m_1, ~m_2, ~m_3, ~\tan\beta, ~\rm {sgn}(\mu)$,
with $m_0$ the universal scalar mass, $A_0$ the universal trilinear coupling, $m_1$, $m_2$, $m_3$ are the $U(1)$, $SU(2)$ and $SU(3)$ gaugino masses,  $\tan\beta$ the ratio of the Higgs vevs and sgn$(\mu)$ is the sign of the Higgs mixing parameter appearing in the superpotential, Eq.~(\ref{5}), which is taken to be positive. Using the soft parameters as input  at the GUT scale, the renormalization group equations (RGE) are run down to the electroweak scale using \code{SoftSUSY 4.1.0}~\cite{Allanach:2001kg, Allanach:2016rxd} which generates the weak scale inputs that enter into the calculation of the observables in this analysis. Also, the SM Higgs boson mass is determined at the two-loop level. The high scale input and the computed Higgs boson masses, consistent with a mass of $125\pm 2$ GeV, for several representative benchmark points are presented in Table~\ref{input}. 

\begin{table}[H]
\begin{center}
\begin{tabulary}{0.85\textwidth}{l|CCCCCCC}
\hline\hline\rule{0pt}{3ex}
Model & $m_0$ & $A_0$ & $m_1$ & $m_2$ & $m_3$ & $\tan\beta$ & $h^0$\\
\hline\rule{0pt}{3ex}  
\!\!(a)  & 3974 & $-10412$ & 486 & 388 & 4517 & 39 & 124.5 \\
(b) & 4769 & $-14593$ & 463 & 245 & 3389 & 23 & 123.5 \\  
(c) & 9026 & $-20940$	 & 484 & 280 & 4143 & 14 & 124.3 \\ 
(d) & 3306 & $-9554$ & 351 & 228 & 2799 & 25 & 123.7 \\
(e) & 7004 & $-8825$ & 619 & 427 & 5194 & 31 & 123.5 \\
\hline
\end{tabulary}\end{center}
\caption{Input parameters for the benchmark points used in this analysis along with the calculated Higgs boson $(h^0)$ mass. The high scale boundary conditions are obtained in the non-universal gaugino sector. All masses are in GeV.}
\label{input}
\end{table}

\begin{table}[H]
\begin{center}
\begin{tabular}{|c | c c c | c|}
\cline{2-4}
\nocell{1}& \multicolumn{3}{c}{Model point} & \nocello{1}\\
\hline\hline
 Observable  & (a) & (b) & (c) & Upper limits \\
\hline
$\mathcal{B}(\mu \rightarrow e \gamma)$ & $3.5\times 10^{-13}$  & $5.0\times 10^{-13}$ & $5.6\times 10^{-13}$ & $5.7 \times 10^{-13}$ \\
$\mathcal{B}(\tau \rightarrow \mu \gamma)$ &$4.1\times 10^{-8}$  & $3.4\times 10^{-8}$ & $4.3\times 10^{-8}$& $4.4 \times 10^{-8}$ \\
 $\mathcal{B}(\tau \rightarrow e \gamma)$ & $3.6 \times 10^{-11}$  & $8.2\times10^{-11}$ & $1.2\times 10^{-10}$ & $3.3 \times 10^{-8}$ \\
\hline
$|d_e|$ & $2.4 \times 10^{-29}$ & $4.8 \times 10^{-29}$ & $4.3\times 10^{-29}$ & $9.3 \times 10^{-29}$ \\
 $|d_{\mu}|$ & $2.1\times 10^{-26}$  & $2.1 \times 10^{-26}$& $2.1\times 10^{-26}$ & $1.9 \times 10^{-19}$ \\
 $|d_{\tau}|$ & $2.5\times 10^{-23}$  & $1.4 \times 10^{-22}$& $2.3\times 10^{-22}$ & $1.1 \times 10^{-17}$ \\
\hline
 $|\Delta a_\mu|$ &  $2.3 \times 10^{-11}$  & $7.1\times 10^{-12}$ & $1.2\times 10^{-12}$ &  $(28.8 \pm 7.9)\times 10^{-10}$ \\
  $|\Delta a_e|$ & $5.4\times 10^{-16}$  & $1.6\times 10^{-16}$ & $2.5\times 10^{-17}$ &   $-10.5 (8.1) \times  10^{-13}$ \\
\hline
$R_{\gamma\gamma}$ & 1.07 & 1.13 & 1.03 &  ATLAS/CMS, Eq.~(\ref{Rgglimits})\\
\hline\hline
\end{tabular}
\caption{An exhibition of the branching ratios {$\mathcal{B}(\ell_i \to \ell_j\gamma$)}, electric dipole moments $|d_\alpha|$, anomalous magnetic moments $\Delta a_\alpha$ and the Higgs diphoton decay enhancement $R_{\gamma\gamma}$ for three benchmark points (a), (b) and (c) of Table~\ref{input}. For point (a), $|f_ 3| = 2.9$, $|f_ 4| = 9.3 $, $|f_ 4{}^{\prime\prime}| = 3.5 \times 10^{-3}$, $|f_ 3{}^{\prime\prime}| = 7.9 \times 10^{-4}$, $\tilde{M}_E=700$, $\tilde{M}_{\chi}=37300$ for point (b), $|f_ 3| = 3$, $|f_ 4| = 5$, $|f_ 4{}^{\prime\prime}| = 7 \times 10^{-3}$, $|f_ 3{}^{\prime\prime}| = 7.9 \times 10^{-4}$, $\tilde{M}_E=800$, $\tilde{M}_{\chi}=20500$ and for point (c) $|f_ 3| = 1$, $|f_ 4| = 25 $, $|f_ 4{}^{\prime\prime}| = 5 \times 10^{-3}$, $|f_ 3{}^{\prime\prime}| = 1\times 10^{-3}$, $\tilde{M}_E=700$, $\tilde{M}_{\chi}=18100$. The remaining scalar masses and trilinear couplings are taken to be universal at $m_ 0^{V}=  5 \times 10^4$ and $ |A_ 0^{V}| = 8 \times 10^3$. Also, common for all points: $|f_ 3'|= 1.8 \times 10^{-2}$, $|f_ 4'| = 1.4 \times 10^{-1}$ , $|f_ 5| = 4.5 \times 10^{-8}$, $|f_ 5'| = 3 \times 10^{-8}$, $|f_ 5{}^{\prime\prime }|= 1.2 \times 10^{-8}$, $|h_ 6| = 9.8$, $|h_ 7| = 2.5$, $|h_ 8| = 498$, $\alpha _ {\mu }=\xi _ 1=\xi _ 2=\alpha _ {A_ 0} = \alpha _ {A_{\tilde{\nu }}} = 0, \chi_3 =3.1, \chi_3' =0.2, \chi_3{}^{\prime\prime}= 1.1, \chi _ 4 = 4.7, \chi _4'=4.0, \chi_4{}^{\prime\prime }= 3.9, \chi _ 5=3.6, \chi _ 5'=3.4, \chi _ 5{}^{\prime\prime }=1.3, \chi_6=3.9, \chi_7= 1.7, \chi_8=6.0$, $m_E=m_N=500$, $m_G=400$ and $m^{\nu}_G=340$. EDM is in $e$cm. All masses are in GeV and all phases in rad.}
\label{observables}
\end{center}
\end{table}

Since SUSY contributions involve the exchange of scalars (sleptons and sneutrinos), the input of Table~\ref{input} suggests that such a contribution will be suppressed due to the high scalar masses (being in the several TeV range). Hence, we expect the mirror and fourth sequential generations to have a more significant contribution to the observables. 
The parameters in the vectorlike sector are chosen so as to be consistent with the lepton masses obtained after diagonalization.
We present in Table~\ref{observables} the results of the observables obtained for three benchmark points, (a), (b) and (c) of Table~\ref{input}. On the right-most column, the experimental limits on the corresponding observables is summarized for comparison purpose and the computed values of the observables 
satisfy these bounds. Thus 
the branching ratios of $\mu\rightarrow e\gamma$ and $\tau\rightarrow\mu\gamma$ are below but  close to their upper limits, especially for points (b) and (c) and could  be probed by  a small 
improvement in experiment. The branching ratio of $\tau\rightarrow e\gamma$ appears to be two to three orders of magnitude smaller than its upper limit. 
However, one can achieve somewhat higher values by varying the Yukawa masses $m_E$ and/or $m_G$ as we will see later.
It is interesting that for the same parameter set the EDM of the electron is also close to its current limit while the EDMs of the muon and of tau are 
five to seven orders of magnitude smaller than the upper limits. The electron and muon anomalous magnetic moments are typically small and the contribution is not significant to explain the $\sim 3 \sigma$ 
deviation if indeed it holds up in improved experiment. 
As for the diphoton rate enhancement there are discernible corrections to the branching ratio but consistent with the current limits from ATLAS and CMS, Eq.~(\ref{Rgglimits}).   Here we note that 
 it was shown in previous works (see, e.g.,\cite{Aboubrahim:2016xuz}) 
  that a muon $g-2$ close to the experimental limit can be obtained via leptonic vectorlike exchange. To see if this is possible with the current 
  constraints 
  we take point (a) from Table~\ref{observables} and modify the input parameters. The results are listed in Table~\ref{observables1} where a muon $g-2$ of $\mathcal{O}(10^{-9})$ and with in the observed $3\sigma$ deviation is obtained. The rest of the observables are still in check but one of the branching ratios, namely, $\tau\rightarrow e \gamma$, has become very small. Also, we have obtained a four orders of magnitude increase in the muon EDM.

\begin{table}[H]
\begin{center}
\begin{tabular}{|c | c | c|}
\hline\hline
 Observable  & Point (a) & Upper limits \\
\hline
$\mathcal{B}(\mu \rightarrow e \gamma)$ & $4.0\times 10^{-14}$ & $5.7 \times 10^{-13}$ \\
$\mathcal{B}(\tau \rightarrow \mu \gamma)$ & $1.3\times 10^{-8}$ & $4.4 \times 10^{-8}$ \\
 $\mathcal{B}(\tau \rightarrow e \gamma)$ & $6.3\times 10^{-23}$ & $3.3 \times 10^{-8}$ \\
\hline
$|d_e|$ & $1.4\times 10^{-36}$ & $9.3 \times 10^{-29}$ \\
 $|d_{\mu}|$ & $2.2\times 10^{-22}$ & $1.9 \times 10^{-19}$ \\
 $|d_{\tau}|$ & $1.4\times 10^{-27}$ & $1.1 \times 10^{-17}$ \\
\hline
 $|\Delta a_\mu|$ & $2.2\times 10^{-9}$ &  $(28.8 \pm 7.9)\times 10^{-10}$ \\
  $|\Delta a_e|$ & $5.4\times 10^{-16}$ &   $-10.5 (8.1) \times  10^{-13}$ \\
\hline
$R_{\gamma\gamma}$ & 1.07 &  ATLAS/CMS, Eq.~(\ref{Rgglimits})\\
\hline\hline
\end{tabular}
\caption{An exhibition of the branching ratios {$\mathcal{B}(\ell_i \to \ell_j\gamma$)}, electric dipole moments $|d_\alpha|$, anomalous magnetic moments $\Delta a_\alpha$ and the Higgs diphoton decay enhancement $R_{\gamma\gamma}$ for the benchmark point (a) of Table~\ref{input}. The input is $|f_ 3| = 0.3$, $|f'_3|=3.8\times 10^2$, $|f''_3|=7.9\times 10^{-6}$, $|f_ 4| = 9.3\times 10^{-4} $, $|f'_4|=3.2\times 10^{-1}$, $|f_ 4{}^{\prime\prime}| = 3.5 \times 10^{-7}$,  $|f_ 5| = 4.5 \times 10^{-8}$, $|f_ 5'| = 3 \times 10^{-8}$, $|f_ 5{}^{\prime\prime }|= 1.2 \times 10^{-8}$, $|h_ 6| = 9.8$, $|h_ 7| = 2.5$, $|h_ 8| = 498$, $\alpha _ {\mu }=\xi _ 1=\xi _ 2=\alpha _ {A_ 0} = \alpha _ {A_{\tilde{\nu }}} = 0, \chi_3 =3.1, \chi_3' =0.2, \chi_3{}^{\prime\prime}= 1.1, \chi _ 4 = 4.7, \chi _4'=4.0, \chi_4{}^{\prime\prime }= 3.9, \chi _ 5=3.6, \chi _ 5'=3.4, \chi _ 5{}^{\prime\prime }=1.3, \chi_6=3.9, \chi_7= 1.7, \chi_8=6.0$, $m_E=m_N=500$, $m_G=400$ and $m^{\nu}_G=340$, $\tilde{M}_E=700$, $\tilde{M}_{\chi}=37300$. The remaining scalar masses and trilinear couplings are taken to be universal at $m_ 0^{V}=  5 \times 10^4$ and $ |A_ 0^{V}| = 8 \times 10^3$. EDM is in $e$cm. All masses are in GeV and all phases in rad.}
\label{observables1}
\end{center}
\end{table}

\begin{figure}[H]
\begin{center}
{\rotatebox{0}{\resizebox*{5.4cm}{!}{\includegraphics{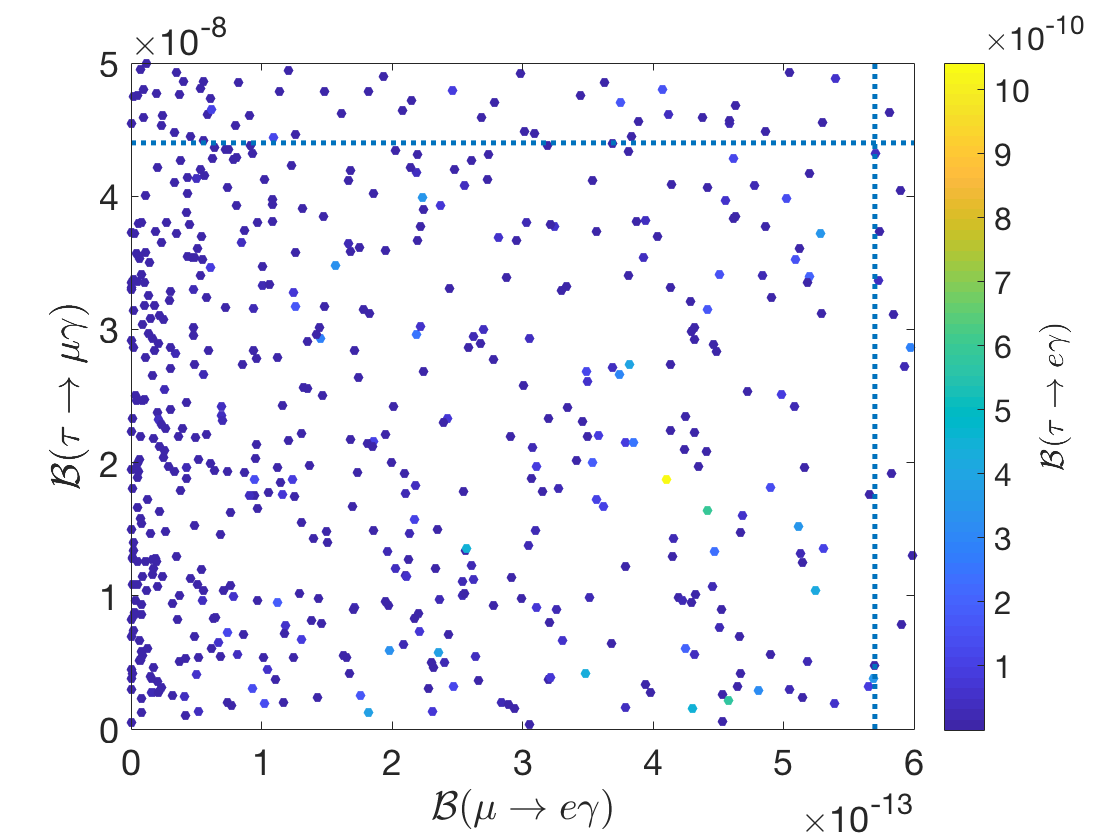}}\hglue5mm}}
{\rotatebox{0}{\resizebox*{5.4cm}{!}{\includegraphics{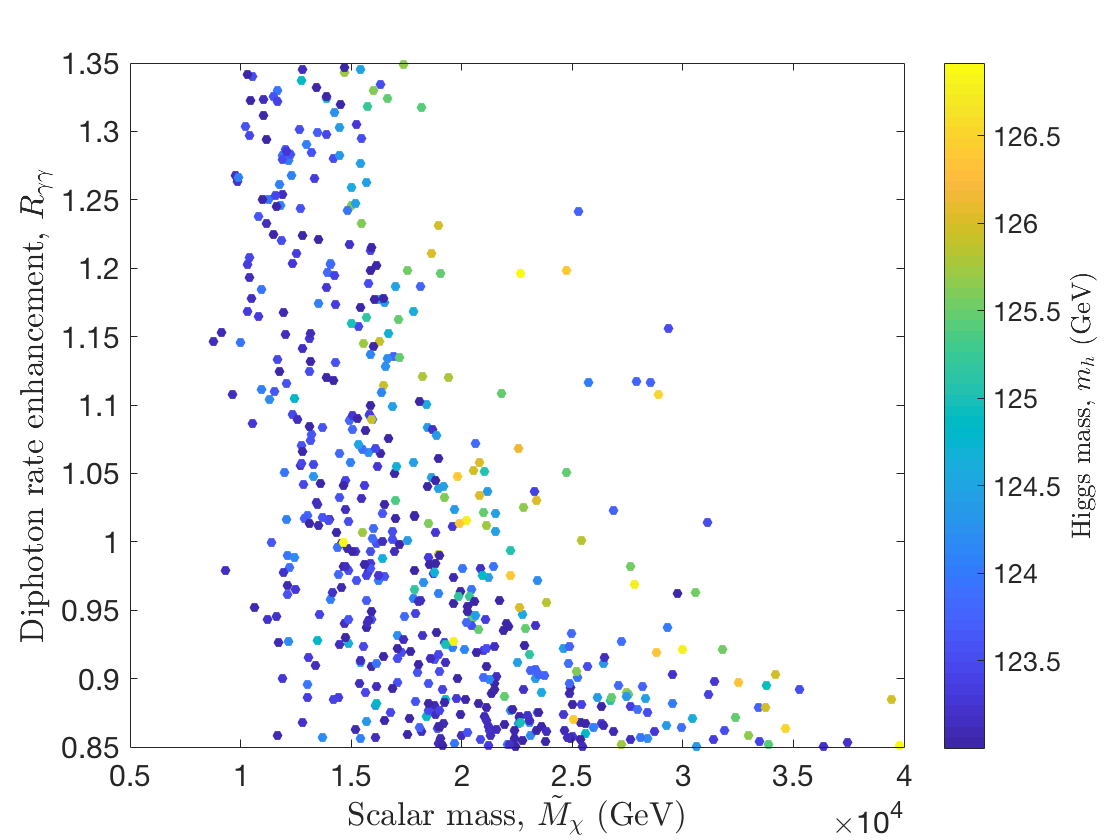}}\hglue5mm}}
{\rotatebox{0}{\resizebox*{5.4cm}{!}{\includegraphics{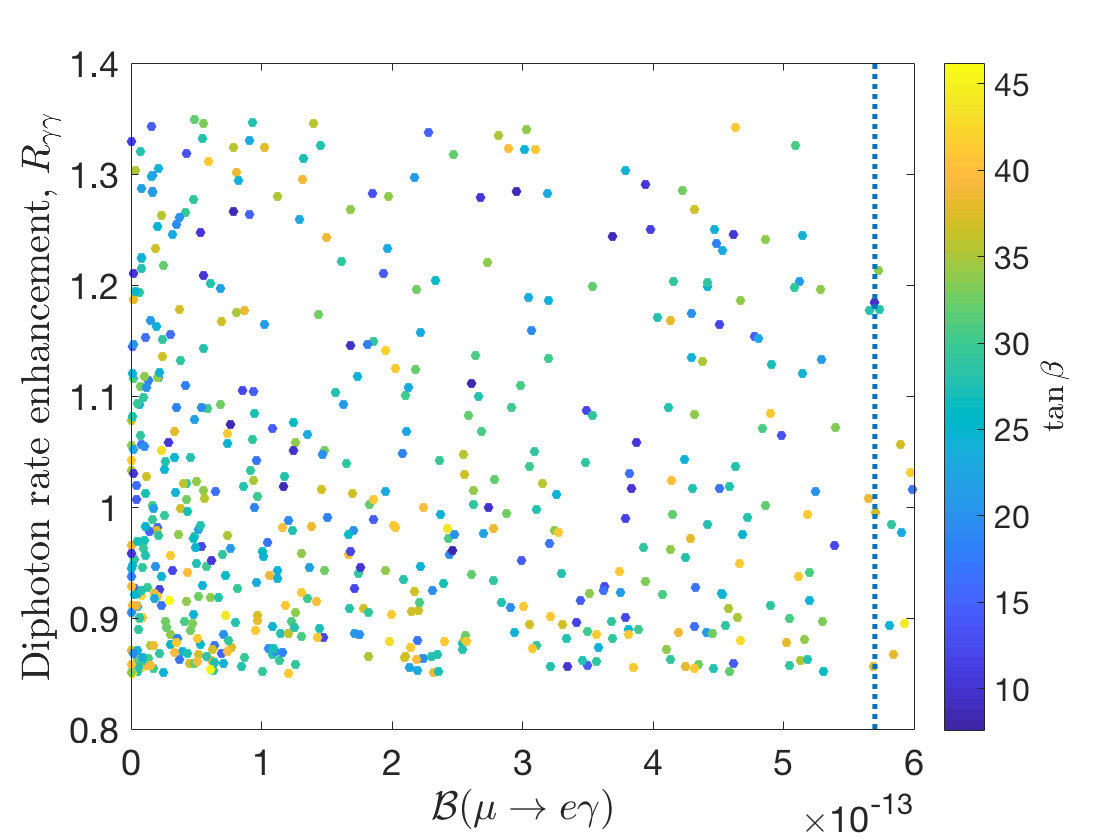}}\hglue5mm}}
\caption{Scatter plots for different observables where the scan is performed over all couplings and their phases for a set of SUGRA benchmark points satisfying the Higgs mass. The upper left panel shows all branching ratios, the upper right panel displays the diphoton enhancement factor, $R_{\gamma\gamma}$ for different values of the scalar mass, $\tilde M_{\chi}$, and the Higgs mass. In the bottom panel, a display of $R_{\gamma\gamma}$ and $\mathcal{B}(\mu\rightarrow e\gamma)$ for different $\tan\beta$. Dashed vertical and horizontal lines correspond to experimental upper limits on the corresponding observables.}
\label{fig5}
\end{center}
\end{figure}

While the results presented here are for an explicit sample set, we have analyzed the parameter space of the model much more widely in the ranges displayed in Eq. (\ref{range}),
\begin{align}
|f_3|\in [1,10], ~~~~& |f'_3|\in [1\times 10^{-2},10], ~~~~|f''_3|\in [1\times 10^{-6},1\times 10^{-4}],  \nonumber \\
|f_4|\in [1,20], ~~~~& |f'_4|\in [1\times 10^{-2},10], ~~~~|f''_3|\in [1\times 10^{-5},1\times 10^{-2}],  \nonumber \\
|h_6|\in [1,20], ~~~~& |h_7|\in [1\times 10^{-3},10], ~~~~|h_8|\in [5,600],  \nonumber \\
\tilde M_E\in [600,800], ~~~~& \tilde M_{\chi}\in [3,5]\times 10^4, ~~~~\chi_i\in [0, 2\pi],
\label{range}
\end{align}
where the vectorlike Yukawa masses are fixed so that  $m_E=m_N=500$ GeV, $m_G=400$ GeV and $m^{\nu}_G=340$ GeV. The couplings $|f_5|$, $|f'_5|$ and $|f''_5|$ are kept small, i.e. $\mathcal{O}(10^{-8})$. The scan results in 17 million points but is greatly reduced when the constraints on the nine observables are applied. The results are displayed as scatter plots in Fig.~\ref{fig5}.\\

\begin{figure}[H]
\begin{center}
{\rotatebox{0}{\resizebox*{6.5cm}{!}{\includegraphics{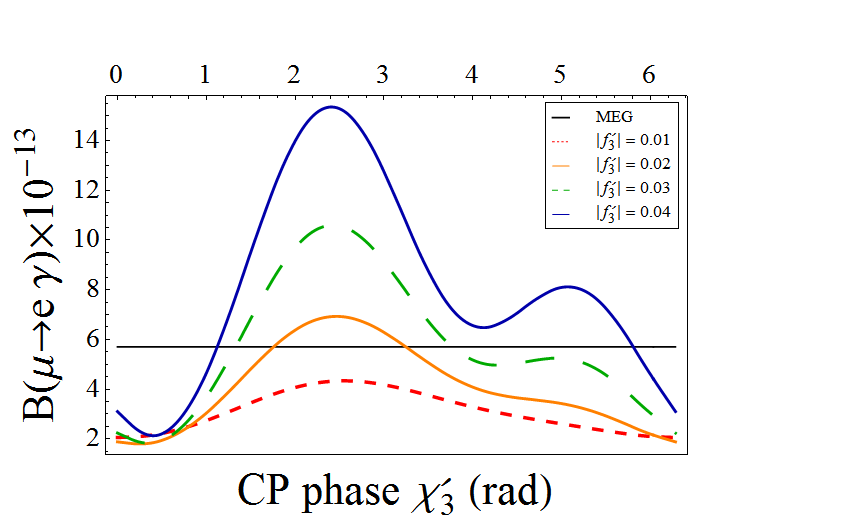}}\hglue5mm}}
{\rotatebox{0}{\resizebox*{6.5cm}{!}{\includegraphics{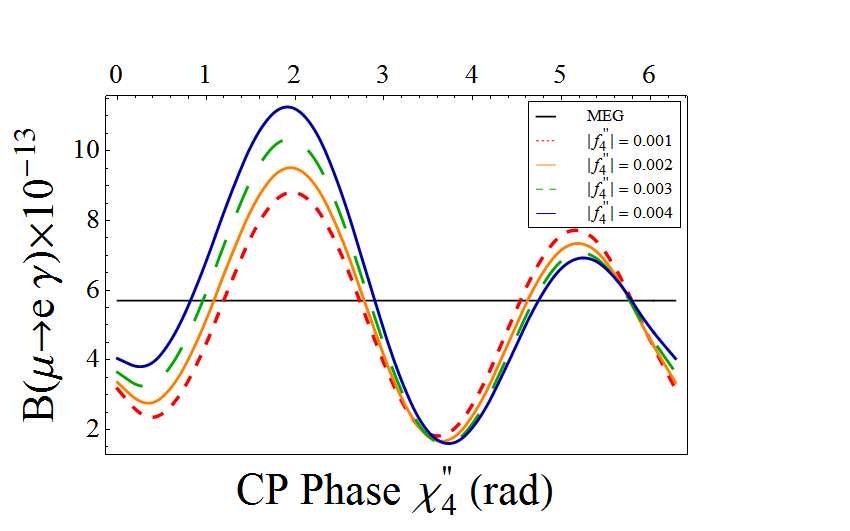}}\hglue5mm}}
{\rotatebox{0}{\resizebox*{6.5cm}{!}{\includegraphics{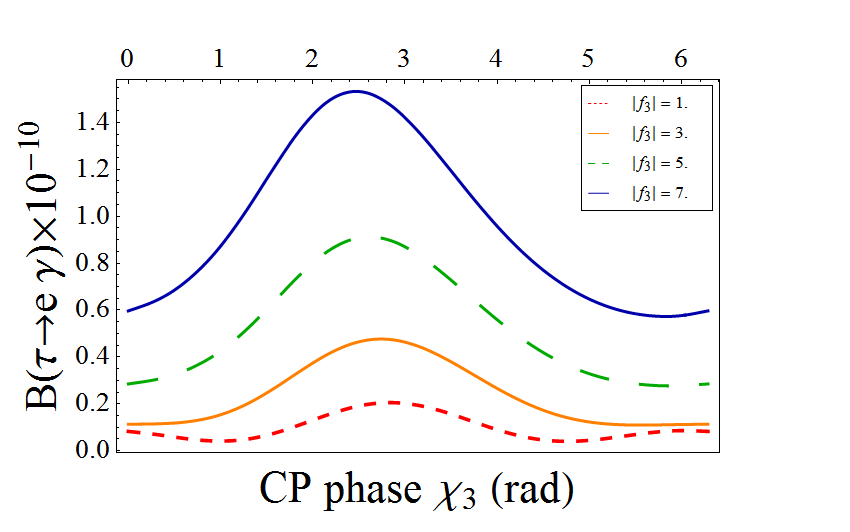}}\hglue5mm}}
{\rotatebox{0}{\resizebox*{6.5cm}{!}{\includegraphics{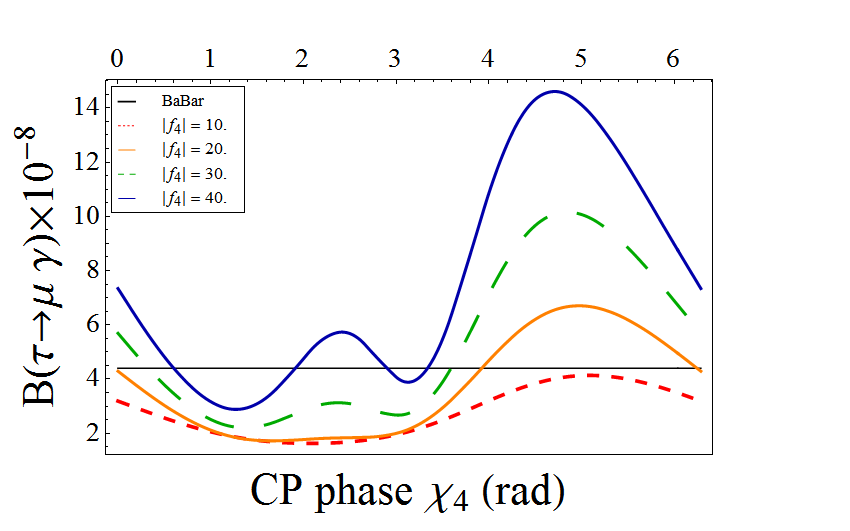}}\hglue5mm}}
{\rotatebox{0}{\resizebox*{6.5cm}{!}{\includegraphics{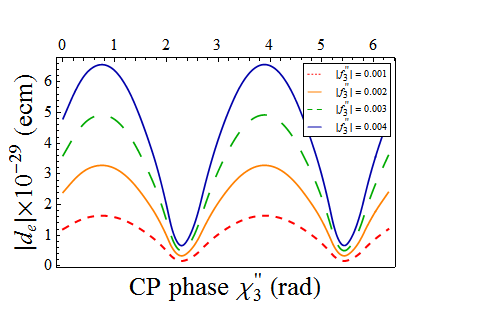}}\hglue5mm}}
\caption{An exhibition of the branching ratios and electron EDM versus the CP phases for point (a) of Table~\ref{input}. The upper panels show $\mathcal{B}(\mu\rightarrow e\gamma)$ as a function of the CP phases $\chi'_3$ and $\chi''_4$ for different values of $|f'_3|$ and $|f''_4|$, respectively. The middle panels show $\mathcal{B}(\tau\rightarrow e\gamma)$ and $\mathcal{B}(\tau\rightarrow \mu\gamma)$ as a function of the CP phases $\chi_3$ and $\chi_4$ for different values of $|f_3|$ and $|f_4|$, respectively. The bottom panel displays the electron EDM versus $\chi''_3$ for different values of $|f''_3|$. All other parameters are the same as for point (a) in Table~\ref{observables}.}
\label{fig6}
\end{center}
\end{figure}

Thus, in the upper left panel of Fig.~\ref{fig5} we display a scatter plot in the three observables, $\mathcal{B}(\mu\rightarrow e\gamma)$, $\mathcal{B}(\tau\rightarrow \mu\gamma)$ and $\mathcal{B}(\tau\rightarrow e\gamma)$. The dashed vertical and horizontal lines are the upper limits on $\mathcal{B}(\mu\rightarrow e\gamma)$ and $\mathcal{B}(\tau\rightarrow \mu\gamma)$, respectively. One can see that there are plenty of points below but close to the upper limits while satisfying all the other observables. The upper right and bottom panels show scatter plots in $R_{\gamma\gamma}$ versus the scalar mass from the vectorlike sector, $\tilde M_{\chi}$ in one and $\mathcal{B}(\mu\rightarrow e\gamma)$ in the other. The Higgs boson mass and $\tan\beta$ are also shown in the $z$-direction. One can see that values of $R_{\gamma\gamma}$ within the experimental limits are more favorable for lower $\tilde M_{\chi}$ values. The reason for this is the following: For vectorlike masses much smaller than their Yukawa counterparts, i.e. $|h_6||h_7|\ll \frac{1}{2}f'_2 y_5 v_1 v_2$, the fermionic contribution, $r_f$ in Eq.~(\ref{rf}), to the diphoton rate enhancement is negative and large ($\sim -0.4$ for the parameter space under consideration). To get values of $R_{\gamma\gamma}$ consistent with experiment, a positive and large contribution must come from the bosonic part, $r_b$, Eq.~(\ref{rb}). It is shown that smaller values of $\tilde M_E$ and $\tilde M_{\chi}$, in the range given by Eq.~(\ref{range}), can achieve this purpose with out affecting other observables. Having this range of values means lighter vectorlike superpartners and the loop contributions become less suppressed.  Since the SUSY loops are suppressed,  the vectorlike sector is the largest contributor to the various observables considered here.

We discuss now in further detail the  sensitivity of some of the observables on the various input parameters.
Thus in Fig.~\ref{fig6} we display the variation of $\mathcal{B}(\mu\rightarrow e\gamma)$, $\mathcal{B}(\tau\rightarrow e\gamma)$, $\mathcal{B}(\tau\rightarrow \mu\gamma)$ and the electron EDM, $|d_e|$ as a function of the CP phases from the vectorlike sector. It is clear that all those observables exhibit a sensitive dependence on the CP phases where the branching ratios oscillate above and below their upper limits. Also, the electron EDM shows large variations very close to the experimental upper limit. The different curves in each plot correspond to different choices of the couplings $|f_3|$, $|f'_3|$, $|f''_3|$, $|f_4|$ and $|f''_4|$ where larger values of the observables are obtained for larger couplings. Note that those couplings cannot take arbitrarly large values since this will spoil the lepton masses.    

In Fig.~\ref{fig7} we show the dependence of the branching ratios of $\mu \rightarrow e\gamma$, $\tau \rightarrow \mu \gamma$, $\tau \rightarrow e \gamma$ and the electron EDM on the vectorlike Yukawa masses for points (c) and (e) of Table~\ref{input}. The observables show a decaying trend for larger values of the masses which is due to larger suppression of loop effects due to the exchange of heavier particles. For point (c), the branching ratio of $\mu \rightarrow e\gamma$ drops below its upper limit for a mass $\sim 250$ GeV while $\tau \rightarrow \mu \gamma$ does that for a heavier mass, $\sim 450$ GeV (top panel). The different curves in each plot correspond to different choices of the vectorlike mass $|h_6|$ where, as one would expect, the contribution from the vectorlike sector is larger for smaller values of $|h_6|$. The interesting aspect of point (e) is in the variation of the branching ratios (middle-right and bottom panels) against $m_E$. As Fig.~\ref{fig7} shows, one can simultaneously get all three branching ratios just below their upper limits by choosing particular values of $m_E=m_G$ and $|h_7|$. Thus, $\mathcal{B}(\tau\rightarrow e\gamma)$ and $\mathcal{B}(\tau\rightarrow \mu\gamma)$ plunge right below their upper limits at around 350 GeV while $\mathcal{B}(\mu\rightarrow e\gamma)$ is already below the upper limit for even smaller $m_E=m_G$. This shows how the interplay of those parameters lead to all constraints to fall in place. While it was difficult to achieve larger $\mathcal{B}(\tau\rightarrow e\gamma)$ values (in Table~\ref{observables}), it was easier to do so for point (e). 

The coupling $f_3$ mixes the vectorlike generation with the first leptonic generation of Eq.~(\ref{8}). Thus we expect this coupling to have the largest impact on observables pertaining to the $\tau$ lepton. To check this, we exhibit the variation of the radiative tau decay branching ratios, the muon EDM and tau EDM against $|f_3|$ in Fig.~\ref{fig8}. The plots are drawn for different values of $|f_4|$. As one can clearly see, the branching ratios of tau and the tau EDM are impacted the most where the former observables may shoot above their upper limits for higher values of $|f_3|$, while the variation of the muon EDM is rather mild. Larger values of $|f_4|$, which couples the vectorlike and first generation singlet fields, produces larger values of the considered observables as one would expect as well.  

\begin{figure}[H]
\begin{center}
{\rotatebox{0}{\resizebox*{6.5cm}{!}{\includegraphics{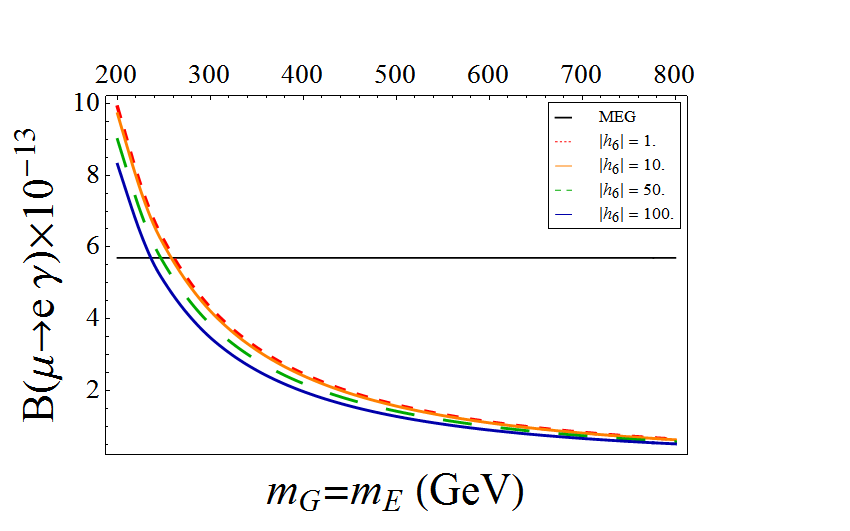}}\hglue5mm}}
{\rotatebox{0}{\resizebox*{6.5cm}{!}{\includegraphics{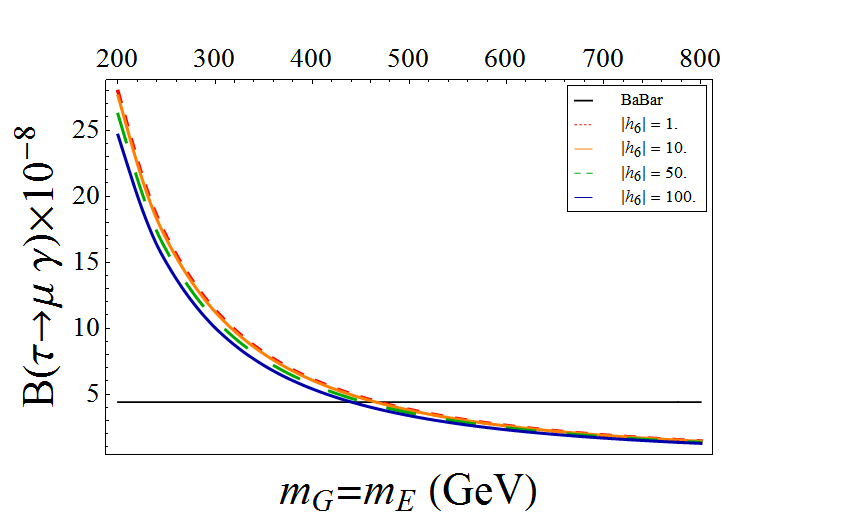}}\hglue5mm}}
{\rotatebox{0}{\resizebox*{6.5cm}{!}{\includegraphics{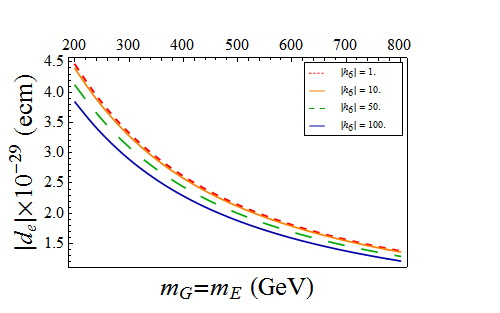}}\hglue5mm}}
{\rotatebox{0}{\resizebox*{6.7cm}{!}{\includegraphics{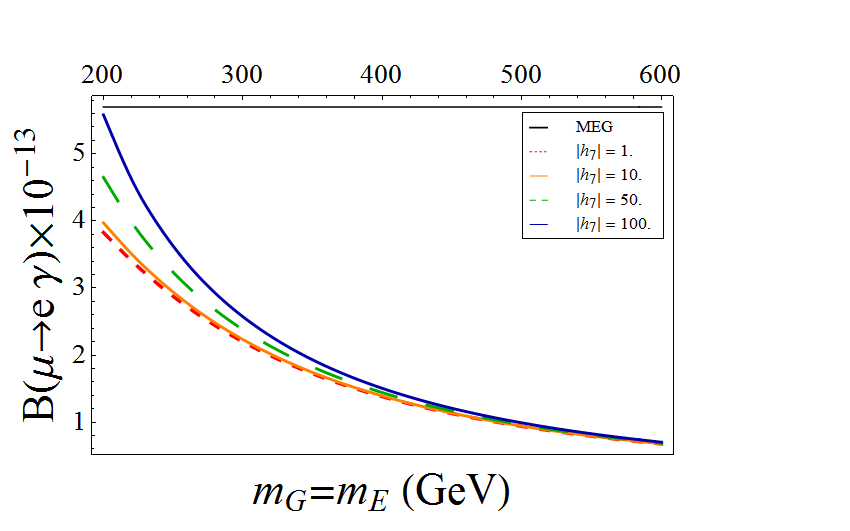}}\hglue5mm}}
{\rotatebox{0}{\resizebox*{6.5cm}{!}{\includegraphics{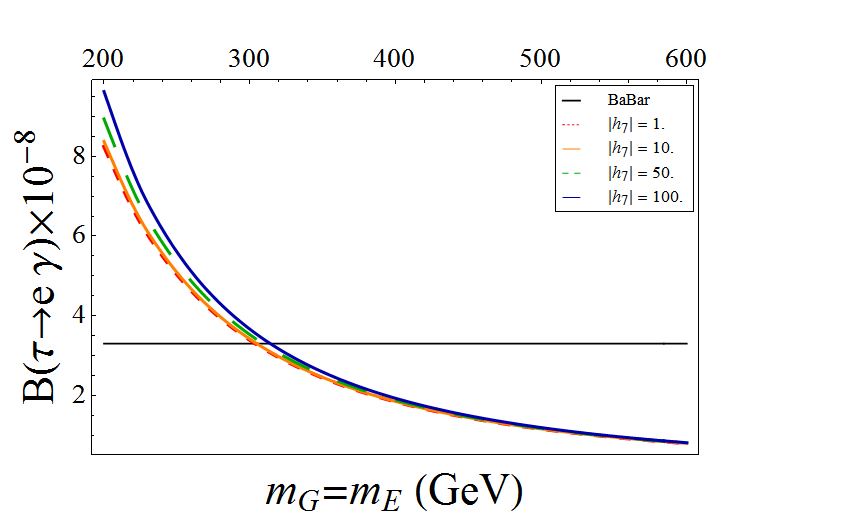}}\hglue5mm}}
{\rotatebox{0}{\resizebox*{6.5cm}{!}{\includegraphics{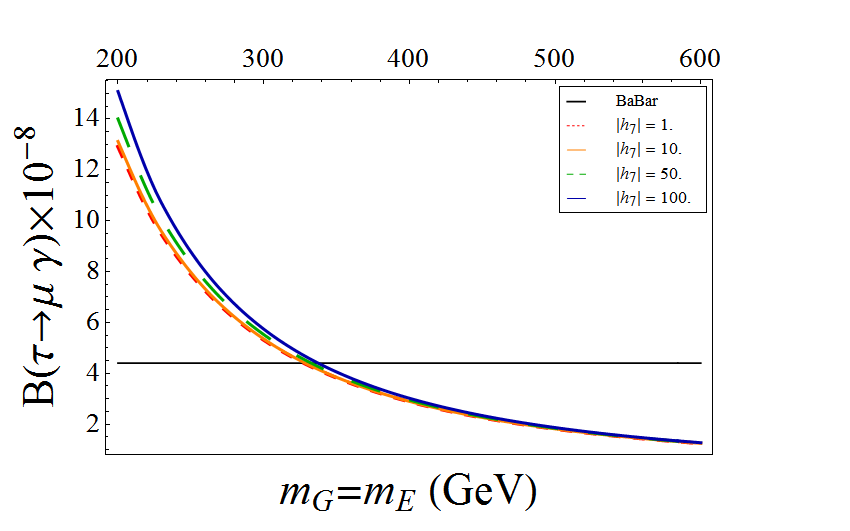}}\hglue5mm}}
\caption{An exhibition of the $\mu \rightarrow e\gamma$ and $\tau \rightarrow \mu \gamma$ branching ratios (top panel) and the electron EDM (middle-left panel) as a function of $m_E(=m_G)$ for different values of $|h_6|$ for benchmark point (c) while plots of all branching ratios (middle-right and bottom panels) are drawn for different $|h_7|$ for point (e) of Table~\ref{input}. For point (c), all other parameters are the same as in Table~\ref{observables}. As for point (e), the other scalar masses are $m^{V}_0=5\times 10^4$ except for $\tilde M_E=700$ and $\tilde M_{\chi}=18100$ and the rest of the trilinear couplings are $|A^{V}_0|=8\times 10^3$. Also, $|f_3|=7$, $|f'_3|=2\times 10^{-3}$, $|f''_3|=2\times 10^{-5}$, $|f_4|=4\times 10^{-1}$, $|f'_4|=5\times 10^{-2}$, $|f''_4|=4\times 10^{-2}$, $|f_5|=4.5\times 10^{-10}$, $|f'_5|=3\times 10^{-10}$, $|f''_5|=1.2\times 10^{-10}$, $|h_6|=9.8$, $|h_8|=4.98\times 10^{2}$, $M_N=500$, $m^{\nu}_G=340$, $\chi_3=3.1$, $\chi'_3=0.2$, $\chi''_3=1.1$, $\chi_4=-1.58$, $\chi'_4=-2.3$, $\chi''_4=-2.35$, $\chi_5=-2.7$, $\chi'_5=-2.9$, $\chi''_5=1.3$, $\chi_6=-2.4$, $\chi_7=1.7$ and $\chi_8=-0.3$. All masses are in GeV and phases in rad.} 
\label{fig7}
\end{center}
\end{figure}

\begin{figure}[H]
\begin{center}
{\rotatebox{0}{\resizebox*{6.5cm}{!}{\includegraphics{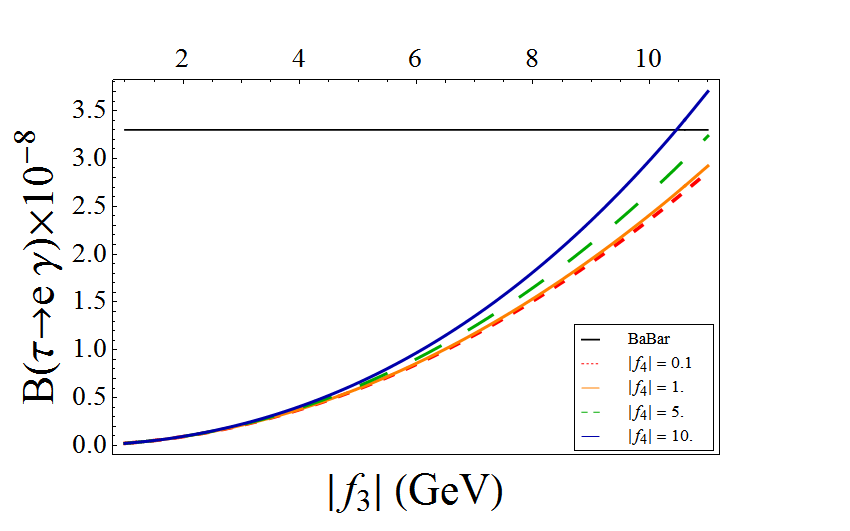}}\hglue5mm}}
{\rotatebox{0}{\resizebox*{6.7cm}{!}{\includegraphics{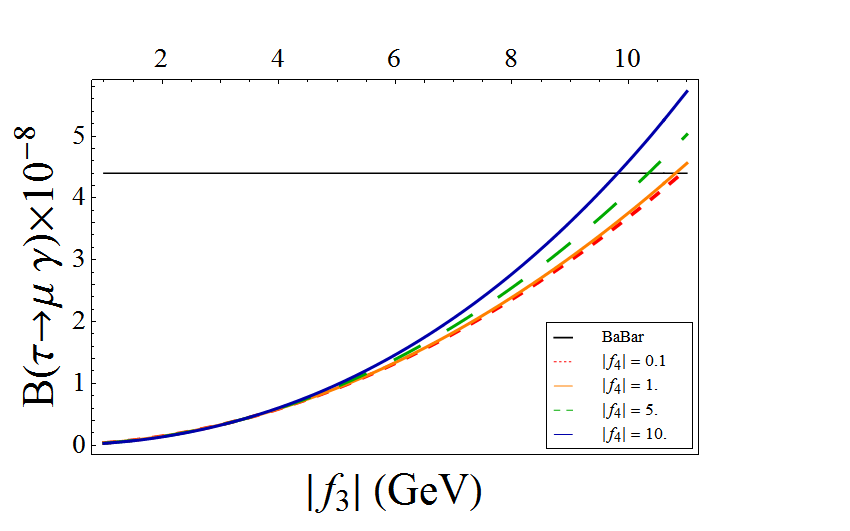}}\hglue5mm}}
{\rotatebox{0}{\resizebox*{6.5cm}{!}{\includegraphics{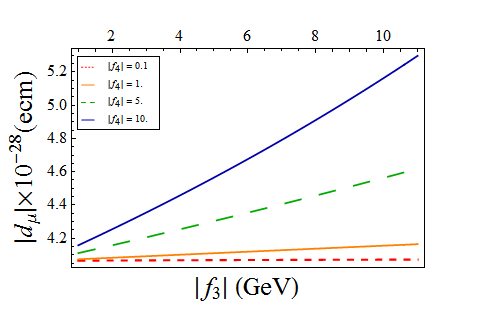}}\hglue5mm}}
{\rotatebox{0}{\resizebox*{6.7cm}{!}{\includegraphics{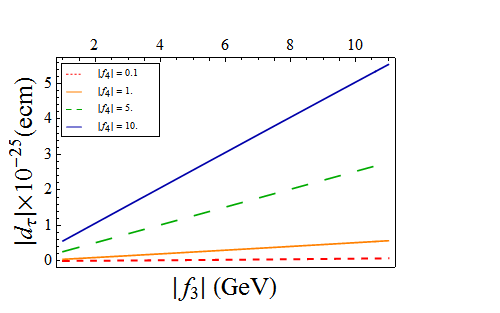}}\hglue5mm}}
\caption{An exhibition of the $\tau \rightarrow e \gamma$ and $\tau \rightarrow \mu \gamma$ branching ratios and the muon and tau EDMs as a function of $|f_3|$ for different values of $|f_4|$ for point (d) of Table~\ref{input}. The other scalar masses are $m^V_0=5\times 10^4$ except for $\tilde M_E=700$ and $\tilde M_{\chi}=18100$ and the rest of the trilinear couplings are $|A^{V}_0|=8\times 10^3$. Also, $|f'_3|=2\times 10^{-3}$, $|f''_3|=2\times 10^{-5}$, $|f'_4|=5\times 10^{-2}$, $|f''_4|=4\times 10^{-2}$, $|f_5|=4.5\times 10^{-10}$, $|f'_5|=3\times 10^{-10}$, $|f''_5|=1.2\times 10^{-10}$, $|h_6|=9.8$, $|h_7|=2.5$, $|h_8|=4.98\times 10^{2}$, $M_N=m_E=500$, $m_G=400$, $m^{\nu}_G=340$, $\chi_3=3.1$, $\chi'_3=0.2$, $\chi''_3=1.1$, $\chi_4=-1.58$, $\chi'_4=-2.3$, $\chi''_4=-2.35$, $\chi_5=-2.7$, $\chi'_5=-2.9$, $\chi''_5=1.3$, $\chi_6=-2.4$, $\chi_7=1.7$ and $\chi_8=-0.3$. All masses are in GeV and phases in rad.} 
\label{fig8}
\end{center}
\end{figure}
 In the above we discussed the lepton flavor changing process $\mu\to e\gamma$ but did not discuss the flavor changing processes
 $\mu\to e$ conversion and $\mu\to 3e$.  A proper treatment of these processes at the same level of care as 
 done for the other processes treated here  is outside the scope of this work. Thus, for example, for the $\mu\to e$ conversion one needs 
  computation of a set of box and penguin diagrams which would again involve our $10\times 10$ scalar mass matrices in the loops. 
  In addition $\mu\to e$ conversion has much more model dependence because of nuclear physics effects. Here we give 
  approximate results for them valid in certain limits which, however, do indicate the expected size of the branching ratios for 
  these processes for the parameter sets in our case given in Table 2.  Thus in the dipole dominance approximation, 
  one has~\cite{Altmannshofer:2013lfa}  
     \begin{align}
  \frac{\mathcal{B}(\mu\to 3e)}{\mathcal{B}(\mu\to e\gamma)} \simeq \frac{\alpha_{em}}{3\pi}  \left (\log\frac{m_{\mu}^2}{m_e^2}- \frac{11}{4}\right).
  \label{mu3e}
  \end{align}   
  The right hand side of Eq.~(\ref{mu3e}) evaluates to $\sim 6\times 10^{-3}$. Using this ratio 
  $\mathcal{B}(\mu\to 3e) \sim 3.4\times 10^{-15}$ for column 3 in Table 2. This is to be compared with the current
  experimental limit~\cite{Bellgardt:1987du}

   \begin{align}
 \mathcal{B}(\mu\to 3e) < 1.0 \times 10^{-12} ~~~ \rm {at} ~90\% ~{\rm C.L.}
  \end{align}  
  In future experiment~\cite{Baldini:2013ke} this limits may reach $\mathcal{B}(\mu\to 3e) \leq 10^{-16}$.
  For the $\mu\to e$ conversion process the analysis of~\cite{Ellis:2016yje} 
   in the limit $m_{\chi^{\pm}}^2/m^2_{\tilde \ell}\sim 1$ gives
     $\mathcal{B}(\mu\to e)_N/ \mathcal{B}(\mu\to e\gamma) \sim \alpha_{em}/3$ for Aluminum and 
      $\mathcal{B}(\mu\to e)_N/ \mathcal{B}(\mu\to e\gamma) \sim \alpha_{em}/2$ for Gold. Numerically, for Aluminum this gives 
       $\mathcal{B}(\mu\to e)_{\rm Al} \sim 1.3 \times 10^{-15}$ and for Gold it gives 
       $\mathcal{B}(\mu\to e)_{\rm G} \sim 1.95 \times 10^{-15}$         
       for the third column in Table 2.   
      The current experimental limit for $\mu\to e$ conversion for Gold is~\cite{Bertl:2006up}
      \begin{align}
     \mathcal{B}(\mu\to e)_{\rm Au} < 7\times 10^{-13}  ~~\rm {at} ~90\% ~{\rm C.L.}.
      \end{align}
      In the future one expects that experiments using Al nuclei will reach a sensitivity in the 
      range~\cite{Hewett:2012ns,Abrams:2012er}   
      $\mathcal{B}(\mu\to e)_{\rm Al}  <  10^{-16} - 10^{-18}$.


\section{Conclusion\label{sec:conclu}}

In a large class of models such as based on grand unification, on strings and branes,  one has vectorlike states some of 
which could be light and lie in the low energy region accessible to experiment. Their presence can affect low energy phenomena
through loop corrections.
 In supersymmetric theories the vectorlike generations will have particles and their mirrors as well
as sparticles and their mirrors. This means that in a model with three generations there will be two more particles that can
appear in the mixing matrix, making the fermionic mixing matrix a  $5\times 5$ mixing matrix. In the slepton sector, one 
will have in general a $10\times 10$ mixing matrix. The analysis is done including the CP violating phases in the mixings of the vectorlike generation.
 In his work we have carried out a correlated study of the effects of the vectorlike generation on several observables. These include  $\mu\to e \gamma$, $\tau \to \mu \gamma$, $\tau\to e\gamma$, 
muon and electron magnetic moments, $g_\mu-2$ and $g_e-2$, and EDMs of the charged leptons $d_e, d_\mu, d_\tau$.
We also examine the effect of the vectorlike generation on $h\to \gamma\gamma$. The analysis is done under the constraints
of the Higgs boson mass at $125$ GeV. Several interesting correlations are observed which are discussed in the
numerical section. In the coming years improvement in experiment on several fronts will occur and the predictions 
of the vectorlike generations can be checked or the model further constrained. \\

\noindent
{\em Acknowledgments:}\
This research was supported in part  by  NSF Grant PHY-1620526.\\


\newpage

\begin{appendices}
\section{The extended MSSM with a vectorlike leptonic generation \label{sec:appendA}}
The mass-squared matrices of the supersymmetric scalar sectors (sleptons, sneutrinos and their mirrors) arise from the $F$ and $D$ terms of the potential and from the soft SUSY breaking terms such that ${\cal L}_{\rm soft}$ takes the form
\begin{align}
-{\cal L}_{\text{soft}}&=\tilde M^2_{\tau L} \tilde \psi^{i*}_{\tau L} \tilde \psi^i_{\tau L}
+\tilde M^2_{\chi} \tilde \chi^{ci*} \tilde \chi^{ci}
+\tilde M^2_{\mu L} \tilde \psi^{i*}_{\mu L} \tilde \psi^i_{\mu L}\nonumber\\
&+\tilde M^2_{e L} \tilde \psi^{i*}_{e L} \tilde \psi^i_{e L}
+\tilde M^2_{\nu_\tau} \tilde \nu^{c*}_{\tau L} \tilde \nu^c_{\tau L}
 +\tilde M^2_{\nu_\mu} \tilde \nu^{c*}_{\mu L} \tilde \nu^c_{\mu L} \nonumber \\
&
+\tilde M^2_{4 L} \tilde \psi^{i*}_{4 L} \tilde \psi^i_{4 L}
+\tilde M^2_{\nu_4} \tilde \nu^{c*}_{4 L} \tilde \nu^c_{4 L}
+\tilde M^2_{\nu_e} \tilde \nu^{c*}_{e L} \tilde \nu^c_{e L}
+\tilde M^2_{\tau} \tilde \tau^{c*}_L \tilde \tau^c_L
+\tilde M^2_{\mu} \tilde \mu^{c*}_L \tilde \mu^c_L\nonumber\\
&
+\tilde M^2_{e} \tilde e^{c*}_L \tilde e^c_L
+\tilde M^2_E \tilde E^*_L \tilde E_L
 + \tilde M^2_N \tilde N^*_L \tilde N_L
+\tilde M^2_{4} \tilde \ell^{c*}_{4L} \tilde \ell^c_{4L}
 \nonumber \\
&+\epsilon_{ij} \{f_1 A_{\tau} H^i_1 \tilde \psi^j_{\tau L} \tilde \tau^c_L
-f'_1 A_{\nu_\tau} H^i_2 \tilde \psi ^j_{\tau L} \tilde \nu^c_{\tau L}
+h_1 A_{\mu} H^i_1 \tilde \psi^j_{\mu L} \tilde \mu^c_L
-h'_1 A_{\nu_\mu} H^i_2 \tilde \psi ^j_{\mu L} \tilde \nu^c_{\mu L} \nonumber \\
&+h_2 A_{e} H^i_1 \tilde \psi^j_{e L} \tilde e^c_L
-h'_2 A_{\nu_e} H^i_2 \tilde \psi ^j_{e L} \tilde \nu^c_{e L}
+f_2 A_N H^i_1 \tilde \chi^{cj} \tilde N_L
-f'_2 A_E H^i_2 \tilde \chi^{cj} \tilde E_L \nonumber\\
&
+y_5 A_{4\ell} H^i_1 \tilde \psi^j_{4 L} \tilde \ell^c_{4L}
-y'_5 A_{4\nu} H^i_2 \tilde \psi ^j_{4 L} \tilde \nu^c_{4 L}
+\text{h.c.}\}.
\label{13}
\end{align}

We define the slepton mass-squared  matrix $M^2_{\tilde \tau}$  in the basis
\beq
 (\tilde  \tau_L, \tilde E_L, \tilde \tau_R,
\tilde E_R, \tilde \mu_L, \tilde \mu_R, \tilde e_L, \tilde e_R, \tilde\ell_{4L}, \tilde\ell_{4R}),
\eeq
and  label the matrix  elements as $(M^2_{\tilde \tau})_{ij}= M^2_{ij}$ where these elements  are given by
\begin{gather}
M^2_{11}=\tilde M^2_{\tau L} +\frac{v^2_1|f_1|^2}{2} +|f_3|^2 -m^2_Z \cos 2 \beta \left(\frac{1}{2}-\sin^2\theta_W\right), \nonumber\\
M^2_{22}=\tilde M^2_E +\frac{v^2_2|f'_2|^2}{2}+|f_4|^2 +|f'_4|^2+|f''_4|^2
+|h_7|^2
 +m^2_Z \cos 2 \beta \sin^2\theta_W, \nonumber\\
M^2_{33}=\tilde M^2_{\tau} +\frac{v^2_1|f_1|^2}{2} +|f_4|^2 -m^2_Z \cos 2 \beta \sin^2\theta_W, \nonumber\\
M^2_{44}=\tilde M^2_{\chi} +\frac{v^2_2|f'_2|^2}{2} +|f_3|^2 +|f'_3|^2+|f''_3|^2
+|h_6|^2
+m^2_Z \cos 2 \beta \left(\frac{1}{2}-\sin^2\theta_W\right), \nonumber\\
M^2_{55}=\tilde M^2_{\mu L} +\frac{v^2_1|h_1|^2}{2} +|f'_3|^2 -m^2_Z \cos 2 \beta \left(\frac{1}{2}-\sin^2\theta_W\right), \nonumber\\
M^2_{66}=\tilde M^2_{\mu} +\frac{v^2_1|h_1|^2}{2}+|f'_4|^2 -m^2_Z \cos 2 \beta \sin^2\theta_W, \nonumber \\
M^2_{77}=\tilde M^2_{e L} +\frac{v^2_1|h_2|^2}{2}+|f''_3|^2 -m^2_Z \cos 2 \beta \left(\frac{1}{2}-\sin^2\theta_W\right), \nonumber\\
M^2_{88}=\tilde M^2_{e} +\frac{v^2_1|h_2|^2}{2}+|f''_4|^2 -m^2_Z \cos 2 \beta \sin^2\theta_W, \nonumber \\
M^2_{99}=\tilde M^2_{4 L} +\frac{v^2_1|y_5|^2}{2} +|h_6|^2 -m^2_Z \cos 2 \beta \left(\frac{1}{2}-\sin^2\theta_W\right), \nonumber\\
M^2_{10 10}=\tilde M^2_{4} +\frac{v^2_1|y_5|^2}{2} +|h_7|^2 -m^2_Z \cos 2 \beta \sin^2\theta_W, \nonumber
\end{gather}
\begin{gather}
M^2_{12}=M^{2*}_{21}=\frac{ v_2 f'_2f^*_3}{\sqrt{2}} +\frac{ v_1 f_4 f^*_1}{\sqrt{2}} ,\nonumber\\
M^2_{13}=M^{2*}_{31}=\frac{f^*_1}{\sqrt{2}}(v_1 A^*_{\tau} -\mu v_2),\nonumber\\
M^2_{14}=M^{2*}_{41}=0, M^2_{15} =M^{2*}_{51}=f'_3 f^*_3,\nonumber \\
 M^{2}_{16}= M^{2*}_{61}=0,  M^{2}_{17}= M^{2*}_{71}=f''_3 f^*_3,  M^{2}_{18}= M^{2*}_{81}=0,
M^2_{23}=M^{2*}_{32}=0,\nonumber\\
M^2_{24}=M^{2*}_{42}=\frac{f'^*_2}{\sqrt{2}}(v_2 A^*_{E} -\mu v_1),  M^2_{25} = M^{2*}_{52}= \frac{ v_2 f'_3f'^*_2}{\sqrt{2}} +\frac{ v_1 h_1 f^*_4}{\sqrt{2}} ,\nonumber\\
 M^2_{26} =M^{2*}_{62}=0,  M^2_{27} =M^{2*}_{72}=  \frac{ v_2 f''_3f'^*_2}{\sqrt{2}} +\frac{ v_1 h_1 f'^*_4}{\sqrt{2}},  M^2_{28} =M^{2*}_{82}=0, \nonumber\\
M^2_{34}=M^{2*}_{43}= \frac{ v_2 f_4 f'^*_2}{\sqrt{2}} +\frac{ v_1 f_1 f^*_3}{\sqrt{2}}, M^2_{35} =M^{2*}_{53} =0, M^2_{36} =M^{2*}_{63}=f_4 f'^*_4,\nonumber\\
 M^2_{37} =M^{2*}_{73} =0,  M^2_{38} =M^{2*}_{83} =f_4 f''^*_4,
M^2_{45}=M^{2*}_{54}=0, M^2_{46}=M^{2*}_{64}=\frac{ v_2 f'_2 f'^*_4}{\sqrt{2}} +\frac{ v_1 f'_3 h^*_1}{\sqrt{2}}, \nonumber\\
 M^2_{47} =M^{2*}_{74}=0,  M^2_{48} =M^{2*}_{84}=  \frac{ v_2 f'_2f''^*_4}{\sqrt{2}} +\frac{ v_1 f''_3 h^*_2}{\sqrt{2}},\nonumber\\
M^2_{56}=M^{2*}_{65}=\frac{h^*_1}{\sqrt{2}}(v_1 A^*_{\mu} -\mu v_2),
 M^2_{57} =M^{2*}_{75}=f''_3 f'^*_3,  M^2_{58} =M^{2*}_{85}=0,  M^2_{67} =M^{2*}_{76}=0,\nonumber\\
 M^2_{68} =M^{2*}_{86}=f'_4 f''^*_4,  M^2_{78}=M^{2*}_{87}=\frac{h^*_2}{\sqrt{2}}(v_1 A^*_{e} -\mu v_2)\nonumber\\
M^{2}_{19}= M^{2*}_{91}=f^*_3 h_6,  M^{2}_{1 10}= M^{2*}_{10 1}=0,\nonumber\\
 M^2_{29} =M^{2*}_{92}=\frac{ v_1 y_5h^*_7}{\sqrt{2}} +\frac{ v_2 h_6 f'^*_2}{\sqrt{2}},  M^2_{2 10} =M^{2*}_{10 2}=0,\nonumber\\
 M^{2}_{39}= M^{2*}_{93}=0,  M^{2}_{3 10}= M^{2*}_{10 3}=f_4 h^*_7,\nonumber\\
 M^2_{49} =M^{2*}_{94}=0,  M^2_{4 10} =M^{2*}_{10 4}=\frac{ v_2 f'_2h^*_7}{\sqrt{2}} +\frac{ v_1 h_6 y^*_5}{\sqrt{2}},\nonumber\\
M^{2}_{59}= M^{2*}_{95}=f'^*_3 h_6,  M^{2}_{5 10}= M^{2*}_{10 5}=0,\nonumber\\
M^{2}_{69}= M^{2*}_{96}=0,  M^{2}_{6 10}= M^{2*}_{10 6}=f'_4 h^*_7,\nonumber\\
M^{2}_{79}= M^{2*}_{97}=f''^*_3 h_6,  M^{2}_{7 10}= M^{2*}_{10 7}=0,\nonumber\\
M^{2}_{89}= M^{2*}_{98}=0,  M^{2}_{8 10}= M^{2*}_{10 8}=f''_5 h^*_7,\nonumber\\
M^2_{9 10}=M^{2*}_{10 9}=\frac{y^*_5}{\sqrt{2}}(v_1 A^*_{4\ell} -\mu v_2).
\label{14}
\end{gather}
   {We assume that  the masses that enter the mass-squared matrix for the scalars are all of electroweak size.
   This mass-squared matrix is hermitian and can be diagonalized with a  unitary transformation} 
\beq
 \tilde D^{\tau \dagger} M^2_{\tilde \tau} \tilde D^{\tau} = \text {diag} (M^2_{\tilde \tau_1},
M^2_{\tilde \tau_2}, M^2_{\tilde \tau_3},  M^2_{\tilde \tau_4},  M^2_{\tilde \tau_5},  M^2_{\tilde \tau_6},  M^2_{\tilde \tau_7},  M^2_{\tilde \tau_8} M^2_{\tilde \tau_9},  M^2_{\tilde \tau_{10}} ).
\label{Dtildetau}
\eeq

The  mass-squared  matrix in the sneutrino sector has a similar structure. In the basis
\beq
(\tilde  \nu_{\tau L}, \tilde N_L,
 \tilde \nu_{\tau R}, \tilde N_R, \tilde  \nu_{\mu L},\tilde \nu_{\mu R}, \tilde \nu_{e L}, \tilde \nu_{e R},
\tilde \nu_{4 L}, \tilde \nu_{4 R}
 ),
\eeq
 {where the sneutrino mass squared matrix
$(M^2_{\tilde\nu})_{ij}=m^2_{ij}$ has elements given by}
\begin{gather}
m^2_{11}=\tilde M^2_{\tau L} +\frac{v^2_2}{2}|f'_1|^2 +|f_3|^2 +\frac{1}{2}m^2_Z \cos 2 \beta,  \nonumber\\
m^2_{22}=\tilde M^2_N +\frac{v^2_1}{2}|f_2|^2 +|f_5|^2 +|f'_5|^2+|f''_5|^2+|h_8|^2, \nonumber\\
m^2_{33}=\tilde M^2_{\nu_\tau} +\frac{v^2_2}{2}|f'_1|^2 +|f_5|^2,  \nonumber\\
m^2_{44}=\tilde M^2_{\chi} +\frac{v^2_1}{2}|f_2|^2 +|f_3|^2 +|f'_3|^2+|f''_3|^2+|h_6|^2 -\frac{1}{2}m^2_Z \cos 2 \beta, \nonumber\\
m^2_{55}=\tilde M^2_{\mu L} +\frac{v^2_2}{2}|h'_1|^2 +|f'_3|^2 +\frac{1}{2}m^2_Z \cos 2 \beta,  \nonumber\\
m^2_{66}=\tilde M^2_{\nu_\mu} +\frac{v^2_2}{2}|h'_1|^2 +|f'_5|^2,  \nonumber\\
m^2_{77}=\tilde M^2_{e L} +\frac{v^2_2}{2}|h'_2|^2+|f''_3|^2+\frac{1}{2}m^2_Z \cos 2 \beta,  \nonumber\\
m^2_{88}=\tilde M^2_{\nu_e} +\frac{v^2_2}{2}|h'_2|^2 +|f''_5|^2,  \nonumber\\
m^2_{99}=\tilde M^2_{4 L} +\frac{v^2_2}{2}|y'_5|^2 +|h_6|^2+\frac{1}{2}m^2_Z \cos 2 \beta,  \nonumber\\
m^2_{10 10}=\tilde M^2_{\nu 4} +|h_8|^2 +\frac{v^2_2}{2}|y'_5|^2 , \nonumber\\
m^2_{12}=m^{2*}_{21}=\frac{v_2 f_5 f'^*_1}{\sqrt{2}}-\frac{ v_1 f_2 f^*_3}{\sqrt{2}},\nonumber\\
m^2_{13}=m^{2*}_{31}=\frac{f'^*_1}{\sqrt{2}}(v_2 A^*_{\nu_\tau} -\mu v_1),
m^2_{14}=m^{2*}_{41}=0,\nonumber\\
m^2_{15}=m^{2*}_{51}= f'_3 f^*_3, m^2_{16}=m^{2*}_{61}=0,\nonumber\\
m^2_{17}=m^{2*}_{71}= f''_3 f^*_3, m^2_{18}=m^{2*}_{81}=0,\nonumber\\
m^2_{23}=m^{2*}_{32}=0,
m^2_{24}=m^{2*}_{42}=\frac{f^*_2}{\sqrt{2}}(v_{1}A^*_N-\mu v_2), m^2_{25}=m^{2*}_{52}=-\frac{v_{1}f^*_2 f'_3}{\sqrt{2}}+\frac{h'_1 v_2 f'^*_5}{\sqrt{2}},\nonumber\\
m^2_{26}=m^{2*}_{62}=0, m^2_{27}=m^{2*}_{72}=-\frac{v_{1}f^*_2 f''_3}{\sqrt{2}}+\frac{h'_2 v_2 f''^*_5}{\sqrt{2}},\nonumber \\
m^2_{28}=m^{2*}_{82}=0, m^2_{34}=m^{2*}_{43}=\frac{v_1 f^*_2 f_5}{\sqrt{2}}-\frac{v_2 f'_1 f^*_3}{\sqrt{2}},\nonumber\\
m^2_{35}=m^{2*}_{53}=0, m^2_{36}=m^{2*}_{63}=f_5 f'^*_5, m^2_{37}=m^{2*}_{73}=0, m^2_{38}=m^{2*}_{83}=f_5 f''^*_5, m^2_{45}=m^{2*}_{54}=0, \nonumber\\
m^2_{46}=m^{2*}_{64}=-\frac{h'^*_1 v_2 f'_3}{\sqrt{2}}+\frac{v_1 f_2 f'^*_5}{\sqrt{2}}, m^2_{47}=m^{2*}_{74}=0, \nonumber\\
m^2_{48}=m^{2*}_{84}=\frac{v_1 f_2 f''^*_5}{\sqrt{2}}-\frac{v_2 h'^*_2 f''_3}{\sqrt{2}}, m^2_{56}=m^{2*}_{65}=\frac{h'^*_1}{\sqrt{2}}(v_2 A^*_{\nu_\mu}-\mu v_1), \nonumber\\
m^2_{57}=m^{2*}_{75}= f''_3 f'^*_3, m^2_{58}=m^{2*}_{85}=0, m^2_{67}=m^{2*}_{76}=0, \nonumber\\
m^2_{68}=m^{2*}_{86}= f'_5 f''^*_5, m^2_{78}=m^{2*}_{87}=\frac{h'^*_2}{\sqrt{2}}(v_2 A^*_{\nu_e}-\mu v_1),\nonumber
\end{gather}

\begin{gather}
m^2_{19}=m^{2*}_{91}= h_6 f^*_3, m^2_{1 10}=m^{2*}_{10 1}=0,\nonumber\\
m^2_{29}=m^{2*}_{92}=-\frac{f_2 v_1 h_6}{\sqrt{2}}+\frac{v_2 h_8 y^*_5}{\sqrt{2}}, m^2_{2 10}=m^{2*}_{10 2}=0, \nonumber\\
m^2_{39}=m^{2*}_{93}=0, m^2_{3 10}=m^{2*}_{10 3}=f_5 h^*_8,\nonumber\\
m^2_{49}=m^{2*}_{94}=0,
m^2_{4 10}=m^{2*}_{10 4}=-\frac{v_2 y'_5 h_6}{\sqrt{2}}+\frac{v_1 h^*_8 f_2}{\sqrt{2}},\nonumber\\
m^2_{59}=m^{2*}_{95}= h_6 f'^*_3, m^2_{5 10}=m^{2*}_{10 5}=0,\nonumber\\
m^2_{69}=m^{2*}_{96}=0, m^2_{6 10}=m^{2*}_{10 6}=f'_5 h^*_8,\nonumber\\
m^2_{79}=m^{2*}_{97}= h_6 f''^*_3, m^2_{7 10}=m^{2*}_{10 7}=0,\nonumber\\
m^2_{89}=m^{2*}_{98}=0, m^2_{8 10}=m^{2*}_{10 8}=f''_5 h^*_8,\nonumber\\
m^2_{9 10}=m^{2*}_{10 9}=\frac{y'_5}{\sqrt{2}}(v_2 A^*_{4 \nu}-\mu v_1).
\label{15}
\end{gather}
  Again as in the charged slepton sector we assume that all the masses are of the electroweak size so all the terms enter in the mass-squared matrix.  This mass-squared  matrix can be diagonalized  by the unitary transformation 
\beq
\tilde D^{\nu\dagger} M^2_{\tilde \nu} \tilde D^{\nu} = \text{diag} (M^2_{\tilde \nu_1}, M^2_{\tilde \nu_2}, M^2_{\tilde \nu_3},  M^2_{\tilde \nu_4},M^2_{\tilde \nu_5},  M^2_{\tilde \nu_6}, M^2_{\tilde \nu_7}, M^2_{\tilde \nu_8},
 M^2_{\tilde \nu_9},  M^2_{\tilde \nu_{10}}
 ).
 \label{Dtildenu}
\eeq



{
\section{Interactions that enter in the analyses of the radiative decays, of the EDMs and of the  magnetic dipole moments of 
the leptons \label{sec:appendB}}
In this appendix we discuss the interactions in the mass diagonal basis involving charged leptons, sneutrinos and charginos. Thus we have
\begin{equation}
-\mathcal{L}_{\tau\tilde{\nu}\tilde{\chi}^{-}}=\sum_{i=1}^2\sum_{j=1}^{10}\bar{\tau}_{\alpha}(C^L_{\alpha i j}P_L+C^R_{\alpha i j}P_R)\tilde{\chi}^{c i}\tilde{\nu}_j+\text{h.c.},
\end{equation}
such that
\begin{align}
\begin{split}
C_{\alpha ij}^{L}=&g(-\kappa_{\tau}U^{*}_{i2}D^{\tau*}_{R1\alpha} \tilde{D}^{\nu}_{1j} -\kappa_{\mu}U^{*}_{i2}D^{\tau*}_{R3\alpha}\tilde{D}^{\nu}_{5j}-
\kappa_{e}U^{*}_{i2}D^{\tau*}_{R4\alpha}\tilde{D}^{\nu}_{7j}\\
&-\kappa_{4\ell}U^{*}_{i2}D^{\tau*}_{R5\alpha}\tilde{D}^{\nu}_{9j}
+U^{*}_{i1}D^{\tau*}_{R2\alpha}\tilde{D}^{\nu}_{4j}-
\kappa_{N}U^{*}_{i2}D^{\tau*}_{R2\alpha}\tilde{D}^{\nu}_{2j})
\end{split} \\~\nonumber\\
\begin{split}
C_{\alpha ij}^{R}=&g(-\kappa_{\nu_{\tau}}V_{i2}D^{\tau*}_{L1\alpha}\tilde{D}^{\nu}_{3j}-\kappa_{\nu_{\mu}}V_{i2}D^{\tau*}_{L3\alpha}\tilde{D}^{\nu}_{6j}-
\kappa_{\nu_{e}}V_{i2}D^{\tau*}_{L4\alpha}\tilde{D}^{\nu}_{8j}+V_{i1}D^{\tau*}_{L1\alpha}\tilde{D}^{\nu}_{1j}+V_{i1}D^{\tau*}_{L3\alpha}\tilde{D}^{\nu}_{5j}\\
&-\kappa_{\nu_{4}}V_{i2}D^{\tau*}_{L5\alpha}\tilde{D}^{\nu}_{10j}
+V_{i1}D^{\tau*}_{L4\alpha}\tilde{D}^{\nu}_{7j}-\kappa_{E}V_{i2}D^{\tau*}_{L2\alpha}\tilde{D}^{\nu}_{4j}),
\end{split}
\end{align}
{where $D_{L,R}^\tau$ and $\tilde D^\nu$ are the charged lepton and sneutrino diagonalizing matrices and 
 are defined by Eq.~(\ref{Dtau}) and Eq.~(\ref{Dtildenu}), respectively and $U$ and $V$ are the matrices that diagonalize the
 chargino mass matrix $M_C$  so that~\cite{Ibrahim:2007fb} 
 \beq
  U^* M_C V^{-1} = \text{diag}(m_{\chi_1^{\pm}} m_{\chi_2^{\pm}})\,.
 \eeq 
Further, 
}
\begin{align}
(\kappa_{N},\kappa_{\tau},\kappa_{\mu},\kappa_{e},\kappa_{4\ell})&=\frac{(m_{N},m_{\tau},m_{\mu},m_{e},m_{4\ell})}{\sqrt{2}m_{W}\cos\beta} , \\~\nonumber\\
(\kappa_{E},\kappa_{\nu_{\tau}},\kappa_{\nu_{\mu}},\kappa_{\nu_{e}},\kappa_{\nu_{4}})&=\frac{(m_{E},m_{\nu_{\tau}},m_{\nu_{\mu}},m_{\nu_{e}},m_{\nu_{4}})}{\sqrt{2}m_{W}\sin\beta} .
\end{align}
{where $m_W$ is the mass of the $W$ boson and $\tan\beta= \langle H_2^2\rangle/\langle H_1^1\rangle$ where $H_1, H_2$ are the
two Higgs doublets of MSSM.}\\
We now discuss the interactions in the mass diagonal basis involving charged leptons, sleptons and neutralinos. Thus we have
\begin{equation}
-\mathcal{L}_{\tau\tilde{\tau}\tilde{\chi}^{0}}=\sum_{i=1}^4\sum_{j=1}^{10}\bar{\tau}_{\alpha}(C^{'L}_{\alpha i j}P_L+C^{'R}_{\alpha i j}P_R)\tilde{\chi}^{0}_i\tilde{\tau}_j+\text{h.c.},
\end{equation}
such that
\begin{align}
C_{\alpha ij}^{'L}=&\sqrt{2}(\alpha_{\tau i}D^{\tau *}_{R1\alpha}\tilde{D}^{\tau}_{1j}-\delta_{E i}D^{\tau *}_{R2\alpha}\tilde{D}^{\tau}_{2j}-
\gamma_{\tau i}D^{\tau *}_{R1\alpha}\tilde{D}^{\tau}_{3j}+\beta_{E i}D^{\tau *}_{R2\alpha}\tilde{D}^{\tau}_{4j}
+\alpha_{\mu i}D^{\tau *}_{R3\alpha}\tilde{D}^{\tau}_{5j}-\gamma_{\mu i}D^{\tau *}_{R3\alpha}\tilde{D}^{\tau}_{6j} \nonumber\\
&+\alpha_{e i}D^{\tau *}_{R4\alpha}\tilde{D}^{\tau}_{7j}-\gamma_{e i}D^{\tau *}_{R4\alpha}\tilde{D}^{\tau}_{8j}
+\alpha_{4\ell i}D^{\tau *}_{R5\alpha}\tilde{D}^{\tau}_{9j}-\gamma_{4\ell i}D^{\tau *}_{R5\alpha}\tilde{D}^{\tau}_{10j}
),
\label{CprimeL}
\end{align}
\begin{align}
C_{\alpha ij}^{'R}=&\sqrt{2}(\beta_{\tau i}D^{\tau *}_{L1\alpha}\tilde{D}^{\tau}_{1j}-\gamma_{E i}D^{\tau *}_{L2\alpha}\tilde{D}^{\tau}_{2j}-
\delta_{\tau i}D^{\tau *}_{L1\alpha}\tilde{D}^{\tau}_{3j}+\alpha_{E i}D^{\tau *}_{L2\alpha}\tilde{D}^{\tau}_{4j}
+\beta_{\mu i}D^{\tau *}_{L3\alpha}\tilde{D}^{\tau}_{5j}-\delta_{\mu i}D^{\tau *}_{L3\alpha}\tilde{D}^{\tau}_{6j}      \nonumber\\
&+\beta_{e i}D^{\tau *}_{L4\alpha}\tilde{D}^{\tau}_{7j}-\delta_{e i}D^{\tau *}_{L4\alpha}\tilde{D}^{\tau}_{8j}
+\beta_{4\ell i}D^{\tau *}_{L5\alpha}\tilde{D}^{\tau}_{9j}-\delta_{4\ell i}D^{\tau *}_{L5\alpha}\tilde{D}^{\tau}_{10j}
),
\label{CprimeR}
\end{align}
where
\begin{align}
\alpha_{E i}&=\frac{gm_{E}X^{*}_{4i}}{2m_{W}\sin\beta} \ ;  && \beta_{E i}=eX'_{1i}+\frac{g}{\cos\theta_{W}}X'_{2i}\left(\frac{1}{2}-\sin^{2}\theta_{W}\right) \\
\gamma_{E i}&=eX^{'*}_{1i}-\frac{g\sin^{2}\theta_{W}}{\cos\theta_{W}}X^{'*}_{2i} \  ;  && \delta_{E i}=-\frac{gm_{E}X_{4i}}{2m_{W}\sin\beta}
\end{align}
and
\begin{align}
\alpha_{\tau i}&=\frac{gm_{\tau}X_{3i}}{2m_{W}\cos\beta} \ ;  && \alpha_{\mu i}=\frac{gm_{\mu}X_{3i}}{2m_{W}\cos\beta} \ ; && \alpha_{e i}=\frac{gm_{e}X_{3i}}{2m_{W}\cos\beta}
 ; && \alpha_{4\ell i}=\frac{gm_{4\ell}X_{3i}}{2m_{W}\cos\beta},
 \\
\delta_{\tau i}&=-\frac{gm_{\tau}X^{*}_{3i}}{2m_{W}\cos\beta} \ ; && \delta_{\mu i}=-\frac{gm_{\mu}X^{*}_{3i}}{2m_{W}\cos\beta} \ ; && \delta_{e i}=-\frac{gm_{e}X^{*}_{3i}}{2m_{W}\cos\beta}
 ; && \delta_{4\ell i}=-\frac{gm_{4\ell}X^{*}_{3i}}{2m_{W}\cos\beta},
\end{align}
{and where }
\begin{align}
\beta_{\tau i}=\beta_{\mu i}=\beta_{e i}=\beta_{4\ell i}&=-eX^{'*}_{1i}+\frac{g}{\cos\theta_{W}}X^{'*}_{2i}\left(-\frac{1}{2}+\sin^{2}\theta_{W}\right),  \\
\gamma_{\tau i}=\gamma_{\mu i}=\gamma_{e i}=\gamma_{4\ell i}&=-eX'_{1i}+\frac{g\sin^{2}\theta_{W}}{\cos\theta_{W}}X'_{2i}\, .
\end{align}
Here $X'$ are defined by
\begin{align}
X'_{1i}&=X_{1i}\cos\theta_{W}+X_{2i}\sin\theta_{W},  \\
X'_{2i}&=-X_ {1i}\sin\theta_{W}+X_{2i}\cos\theta_{W},
\end{align}
where $X$ diagonalizes the neutralino mass matrix, i.e.,
\beqn
X^{T}M_{\chi^{0}}X=\text{diag}(m_{\chi^{0}_{1}},m_{\chi^{0}_{2}},m_{\chi^{0}_{3}},m_{\chi^{0}_{4}}).
\eeqn
{Further, $\tilde D^\tau$ that enter in Eqs. (\ref{CprimeL}) and (\ref{CprimeR}) 
is a matrix which diagonalizes the charged slepton mass squared matrix 
and is defined in Eq. (\ref{Dtildetau}). }\\
In addition to the supersymmetric loop diagrams, we compute the contributions arising from the exchange of $W$ and $Z$ bosons and leptons and mirror leptons in the loops. For the $W$ boson exchange the interactions are given by
\begin{equation}
-\mathcal{L}_{\tau W\psi}=W^{\dagger}_{\rho} \sum_{i=1}^5\sum_{\alpha=1}^{5}\bar{\psi}_{i}\gamma^{\rho}(C^W_{L i\alpha}P_L+C^W_{R i\alpha}P_R)\tau_{\alpha}+\text{h.c.},
\end{equation}
where 
\beqn
C_{L_{i\alpha}}^W= \frac{g}{\sqrt{2}} [D^{\nu*}_{L1i}D^{\tau}_{L1\alpha}+
D^{\nu*}_{L3i}D^{\tau}_{L3\alpha}+D^{\nu*}_{L4i}D^{\tau}_{L4\alpha}
+D^{\nu*}_{L5i}D^{\tau}_{L5\alpha}],
\eeqn
and
\beqn
C_{R_{i\alpha}}^W= \frac{g}{\sqrt{2}}[D^{\nu*}_{R2i}D^{\tau}_{R2\alpha}].
\eeqn
{Here $D_{L,R}^\nu$ are matrices of  a biunitary transformation that diagonalizes the neutrino mass matrix 
and are defined in Eq. (\ref{Dnu}).}
For the $Z$ boson exchange the interactions that enter are given by
\begin{equation}
-\mathcal{L}_{\tau\tau Z}=Z_{\rho} \sum_{\alpha=1}^5\sum_{\beta=1}^{5}\bar{\tau}_{\alpha}\gamma^{\rho}(C^Z_{L\alpha\beta}P_L+C^Z_{R\alpha\beta}P_R)\tau_{\beta},
\end{equation}
where
\beqn
C_{L_{\alpha \beta}}^Z=\frac{g}{\cos\theta_{W}} [x(D_{L\alpha 1}^{\tau\dag}D_{L1\beta}^{\tau}+D_{L\alpha 2}^{\tau\dag}D_{L2\beta}^{\tau}+D_{L\alpha 3}^{\tau\dag}D_{L3\beta}^{\tau}+D_{L\alpha 4}^{\tau\dag}D_{L4\beta}^{\tau}
+D_{L\alpha 5}^{\tau\dag}D_{L5\beta}^{\tau}
)\nonumber\\
-\frac{1}{2}(D_{L\alpha 1}^{\tau\dag}D_{L1\beta}^{\tau}+D_{L\alpha 3}^{\tau\dag}D_{L3\beta}^{\tau}+D_{L\alpha 4}^{\tau\dag}D_{L4\beta}^{\tau}
+D_{L\alpha 5}^{\tau\dag}D_{L5\beta}^{\tau}
)],
\eeqn
and
\beqn
C_{R_{\alpha \beta}}^Z=\frac{g}{\cos\theta_{W}} [x(D_{R\alpha 1}^{\tau\dag}D_{R1\beta}^{\tau}+D_{R\alpha 2}^{\tau\dag}D_{R2\beta}^{\tau}+D_{R\alpha 3}^{\tau\dag}D_{R3\beta}^{\tau}+D_{R\alpha 4}^{\tau\dag}D_{R4\beta}^{\tau}
+D_{R\alpha 5}^{\tau\dag}D_{R5\beta}^{\tau}
)\nonumber\\
-\frac{1}{2}(D_{R\alpha 2}^{\tau\dag}
D_{R 2\beta }^{\tau}
 )]\,.
\eeqn
with $x=\sin^{2}\theta_{W}$. \\
\\
 
}

\end{appendices}

\end{document}